\documentclass[12pt]{article}
\usepackage{amssymb,amsmath,latexsym}
\title{Random Dirac operators with time reversal symmetry}

\author{Christian Sadel, Hermann Schulz-Baldes
\\
\\
{\small Department Mathematik, Universit\"at Erlangen-N\"urnberg, Germany}
}

\date{ }

\newtheorem{theo}{Theorem}
\newtheorem{defini}{Definition}
\newtheorem{proposi}{Proposition}
\newtheorem{lemma}{Lemma}
\newtheorem{coro}{Corollary}

\newcommand{\CC}{{\mathbb C}}
\newcommand{\NN}{{\mathbb N}}
\newcommand{\RR}{{\mathbb R}}

\newcommand{\TM}{{\mathbb T}}
\newcommand{\ZZ}{{\mathbb Z}}
\newcommand{\LM}{{\mathbb L}}

\newcommand{\GM}{{\mathbb G}}
\newcommand{\UM}{{\mathbb U}}
\newcommand{\WM}{{\mathbb W}}

\newcommand{\Aa}{{\cal A}}
\newcommand{\Pp}{{\mathbb S}}
\newcommand{\PP}{{\bf P}}

\newcommand{\pp}{{\bf p}}
\newcommand{\EE}{{\bf E}}

\newcommand{\Dd}{{\cal D}}

\newcommand{\Ww}{{\cal W}}
\newcommand{\Vv}{{\cal V}}
\newcommand{\Ss}{{\cal S}}
\newcommand{\Uu}{{\cal U}}
\newcommand{\Oo}{{\cal O}}
\newcommand{\Tr}{\mbox{\rm Tr}}
\newcommand{\Tt}{{\cal T}}
\newcommand{\Rr}{{\cal R}}
\newcommand{\Nn}{{\cal N}}
\newcommand{\Mm}{{\cal M}}

\newcommand{\Jj}{{\cal J}}

\newcommand{\Qq}{{\cal Q}}
\newcommand{\Kk}{{\cal K}}

\newcommand{\Yy}{{\cal Y}}
\newcommand{\Xx}{{\cal X}}

\newcommand{\loc}{{\mbox{\rm\tiny loc}}}

\newcommand{\one}{{\bf 1}}

\newcommand{\inv}{{\mbox{\rm\tiny inv}}}

\newcommand{\diag}{{\rm diag}}
\newcommand{\SO}{{\rm SO}}

\newcommand{\VM}{{\mathbb V}}

\begin{document}

\maketitle

\begin{abstract}
Quasi-one-dimensional stochastic Dirac operators with an odd number of
channels, time reversal symmetry but otherwise efficiently coupled
randomness are shown to have one conducting channel and absolutely
continuous spectrum of multiplicity two. This follows by adapting the
criteria of Guivarch-Raugi and Goldsheid-Margulis to the analysis of
random products of matrices in the group SO$^*(2L)$, and then a
version of Kotani theory for these operators. Absence of singular
spectrum can be shown by adapting an argument of Jaksic-Last if the
potential contains random Dirac peaks with absolutely continuous
distribution.  
\end{abstract}

\vspace{.5cm}

\section{Introduction}

In this paper we consider a random family of Dirac operators $H$ on the Hilbert space $L^2(\RR,\CC^{2L})$ of square integrable functions with fibers of dimension $L\in\NN$. It is of the form
\begin{equation}
\label{eq-Dirac}
H\;=\;\Jj\,\partial\;+\;\Ww\;+\;\sum_{j\in\ZZ}\;\Vv_j\,\delta_{x_j},
\qquad
\Jj\;=\;\left(\begin{array}{cc} 0 & -\one \\ \one & 0 \end{array} \right)\;,
\end{equation}
where $\partial$ is the space derivative, the potential $\Ww$ is a locally integrable function with values in the hermitian matrices $\mbox{Her}(2L,\CC)$ of size $2L$ and
$\Vv_j\in\,\mbox{Her}(2L,\CC)$ are singular potentials at the points $x_j\in\RR$ (defined as usual by boundary conditions at $x_j$, see Section~\ref{sec-WT}). The potential $\Ww$
is a particular space-homogeneous random process described in detail below, and the $\Vv_j$ are independent and identically distributed. Both potentials are supposed to satisfy time reversal symmetry
\begin{equation}
\label{eq-TRS}
\Jj^*\overline{\Ww(x)}\,\Jj\;=\;\Ww(x)\;,\qquad
\Jj^*\overline{\Vv_j}\,\Jj\;=\;\Vv_j\;.
\end{equation}
This means that $\Jj \Ww(x)$ and $\Jj \Vv_j$ are elements of the Lie algebra so$^*(2L)$ of the classical Lie group SO$^*(2L)$ given by those complex $2L\times 2L$ matrices $\Tt$ satisfying $\Tt^*\Jj\Tt=\Jj$ and $\Tt^t\Tt=\one$.
Hence the Hamiltonian $H$ is self-dual, namely $\Jj^*\overline{H}\Jj=H$, and in the so-called symplectic symmetry class describing time-reversal invariant particles with odd spin. Apart from this symmetry, we suppose the coupling of the potential to be efficient. This is guaranteed if the distribution of $\Jj \Vv_j$ has an absolutely continuous component w.r.t. the volume measure on so$^*(2L)$, but can also be satisfied by adequate choice of $\Ww$ if the $\Vv_j$'s vanish. A more technical formulation of the (actually much weaker) coupling hypothesis is given below in Section~\ref{sec-TRI}. Our main new result is now:

\begin{theo}
\label{theo-main}
Consider the random Dirac operator {\rm \eqref{eq-Dirac}} with time
reversal invariance {\rm \eqref{eq-TRS}} satisfying the {\rm Coupling
  Hypothesis} on the randomness stated in {\rm
  Section~\ref{sec-random}}. 

\vspace{.1cm}

\noindent {\rm (i)} For even channel number $L$, the spectrum of $H$ is almost
surely singular.

\vspace{.1cm}

\noindent {\rm (ii)} For odd channel number $L$, $H$ has almost surely
absolutely continuous spectrum of multiplicity 

2 on all of $\RR$. If
the distribution of the $\Jj \Vv_j$ is absolutely continuous on
$\mbox{\rm so}^*(2L)$, the absolute-

ly continuous spectrum of $H$ is
almost surely pure. 
\end{theo}

Theorem~\ref{theo-main} does not say anything about the singular
spectrum in general ({\it i.e.} without the supplementary assumption
on the distribution of the $\Vv_j$'s), but we believe it to be always
empty. It is crucial that $L$ is odd, as discussed by several authors
in the physics literature (please consult \cite{EM} for a long list of
relevant references). We believe that for even $L$ the spectrum is
almost surely pure-point, but did not try to prove this in detail (it
should be possible by adapting the techniques of \cite{KLS,Bou}). The
main difference between the odd and even case is that there are two
vanishing Lyapunov exponents in the odd case and no vanishing Lyapunov
exponent in the even case. This is related to Kramers' degeneracy and
symplectic symmetry of the Lyapunov spectrum and is proved in
Section~\ref{sec-Lyap}. Based on this fact, the proof of
Theorem~\ref{theo-main} goes on by applying Kotani theory for Dirac
operators as developed by Sun \cite{Sun} along the lines of the work
by Kotani and Simon \cite{KS}. Even though most of the main identities
in \cite{Sun} are correct, it contains some errors which we felt
necessary to correct here. Section~\ref{sec-Kotani} also generalizes
the works \cite{KS,Sun} to singular and complex-valued
potentials. This extension of Kotani theory is non-trivial and crucial
for two reasons: the Coupling Hypothesis cannot be satisfied for real
potentials (see the arguements below) and the singular potentials are
perturbations of finite rank. The latter leads to
similar formulas for the Green functions as in rank one perturbation
theory. Thus the last claim of the theorem can be proved
by adapting the argument of Jaksic and
Last \cite{JL} (see Section~\ref{sec-Absence}).
Sections~\ref{sec-WT} to \ref{sec-random} contain preparatory
material some of which doesn't seem to have appeared in the literature
and makes this work essentially self-contained. 

\vspace{.1cm}

Let us put Theorem~\ref{theo-main} in some perspective, both from a mathematical point of view and a physical one. Most quasi-one-dimensional discrete and continuous random Schr\"odinger operators exhibit Anderson localization, even though some peculiarities such as in the random polymer model may lead to non-trivial quantum diffusion \cite{JSS}. The situation is different for first order differential operators. For example, consider $h=\one\otimes\imath\partial+v$ on $L^2(\RR,\CC^L)$ where $v\in L^\infty(\RR,\mbox{Her}(L,\CC))$ is an essentially bounded hermitian potential (which may be thought of as random). Then the initial value problem $\partial u=\imath vu$, $u(0)=\one$, has a unique solution $u=u(x)$, which lies in the unitary group U$(L)$. Let us use it to define a unitary $\Uu$ on $L^2(\RR,\CC^L)$ by $(\Uu\psi)(x)=u(x)\psi(x)$. Then $\Uu^*h\,\Uu=\one\otimes\imath\partial$ showing that $h$ has absolutely continuous spectrum of multiplicity $L$ for any potential $v$. In physical terms, the operator $h$ can be thought off as an effective model for the chiral edge states of a quantum Hall system with edge conductivity $L$, and the above shows that the nature of the spectrum is conserved under perturbation by a potential, as is the Landauer conductivity which is equal to $L$ (because $\Uu$ commutes with the position operator $X$ on $L^2(\RR,\CC^L)$). Note that the stability of the nature of the spectrum could also be deduced from Mourre theory because $\imath [h,X]=\one$. For true edge states of a disordered magnetic operator on a half-plane, the proof of conservation of absolutely continuous spectrum \cite{BP,FGW} and the edge conductivity \cite{KRS} is much more involved, but possible.

\vspace{.1cm}

Next let us explain why we believe that Mourre theory cannot be applied to the Dirac operator $H$ because there is no natural conjugation operator. In fact, the only physically reasonable choice would be the spin current given by the time derivative of the self-adjoint observable $A=\imath\Jj X$ where is $X$ is the position operator. However, $\imath [A,\Jj\partial+\Ww]=\one+X\Jj(\Ww-\Jj^*\Ww\Jj)$ is positive only if the time-reversal invariant potential $\Ww$ is real and thus $\Jj\Ww$ is in the Lie algebra so$(2L)$. In this situation the Coupling Hypothesis is not satisfied and all Lyapunov exponents vanish. Theorem~\ref{theo-main} is hence a much more delicate result than the one for $h=\imath\partial+v$ just described. We also find it to be a challenging problem to prove absolute continuity of the spectrum for half-plane models for which \eqref{eq-Dirac} is an effective description of the edge states. 

\vspace{.1cm}

Next let us comment on the physical relevance of the Dirac Hamiltonian \eqref{eq-Dirac} with time reversal invariance \eqref{eq-TRS}. It is believed to be an effective model for so-called helical edge states in graphene sheets with a gap at the Dirac point (opened by spin-orbit coupling \cite{EM}). In such graphene sheets the number of edge channels with spin up and spin down is odd and hence these edge states are protected against localization. This is reflected by Theorem~\ref{theo-main}.

\vspace{.2cm}

\noindent {\bf Acknowledgment:} We thank M. Zirnbauer for raising our interest in the time reversal invariant stochastic Dirac operators and the Newton Institute for hospitality and support during our stay in Cambridge. We also thank the Cambridge Philosophical Society for supporting the stay of Christian Sadel at the Newton Institute. This work was funded by the DFG.

\section{Weyl-Titchmarsh matrices}
\label{sec-WT}

This section introduces and analyzes Weyl-Titchmarsh matrices for a fixed non-random Dirac operator
with point interactions. In part this is review (compare {\it e.g.} \cite{HS}) and therefore proofs are kept short, but results need to be written out if only to fix notations. Let $\Pp=(x_j)_{j\in\ZZ}$ be a discrete subset of $\RR$ with no accumulation point and associate to each so-called singular point $x_j$ a singular potential $\Vv_j\in \mbox{Her}(2L,\CC)$. Furthermore let $\Ww$ be in the space $L^1_\loc(\RR,\mbox{Her}(2L,\CC))$ of locally integrable functions with values in the hermitian matrices of size $2L$. All this data encoded in $\omega=(\Ww,(x_j,\Vv_j)_{j\in\ZZ})$, but in this and the next section $\omega$ is fixed and hence suppressed in all notations. The time-reversal symmetry \eqref{eq-TRS} is implemented only in Section~\ref{sec-TRI}. The first aim is to make mathematical sense out of $H$ given in \eqref{eq-Dirac} as a self-adjoint operator on $L^2(\RR,\CC^{2L})$. As usual, the singular potential is dealt with as a certain self-adjoint extension. Before going on, let us point out that most results of this paper also hold for the self-adjoint operator $\Rr\partial+\Ww$ where $x\mapsto \Rr(x)$ is bounded, invertible, and satisfies $\Rr^*=-\Rr$ as well as $\partial\Rr=\Ww^*-\Ww$. In order to focus on the essential difficulties, we stick to the case $\Rr=\Jj$.

\vspace{.2cm}

Let $W^{1,2}(\RR/\Pp,\CC^{2L})$ be the Sobolev space of functions $L^2(\RR/\Pp,\CC^{2L})$ with square-integrable first distributional derivative. Note that these functions $\psi$ are continuous away from $\Pp$ and have left and right limit values $\psi(x\pm)=\lim_{\epsilon\downarrow 0}\psi(x\pm\epsilon)$ for all $x\in\RR$. First we consider the restriction $H_0=H|_{\Dd(H_0)}$ to the domain
$$
\Dd(H_0)\;=\;\left\{\left.
\psi\in W^{1,2}(\RR/\Pp,\CC^{2L})\,\right|\,\psi(x+)=\psi(x-)=0\,\mbox{ for }x\in\Pp\;
\right\}\;.
$$
Then the domain of the adjoint is $\Dd(H_0^*)=W^{1,2}(\RR/\Pp,\CC^{2L})$. The proof of the following result is adapted from \cite{LM}.

\begin{proposi}
\label{prop-defect} For $\psi,\phi\in\Dd(H_0^*)$, one has
\begin{equation}
\label{eq-defect}
\langle H^*_0\psi\,|\,\phi\rangle\,-\,\langle \psi\,|\,H^*_0\phi\rangle\;=\;
\sum_{x\in \Pp}\;\bigl(\,
\psi(x+)^*\Jj\phi(x+)
\,-\,\psi(x-)^*\Jj\phi(x-)\,
\bigr)
\;,
\end{equation}
where the scalar product on the l.h.s. is in $L^2(\RR,\CC^{2L})$ and those on the r.h.s. in $\CC^{2L}$.
\end{proposi}

\noindent {\bf Proof.} Let $\chi_n\in C^\infty(\RR,[0,1])$ with $\chi_n|_{[-n,n]}=1$, $\chi_n|_{[-2n,2n]^c}=0$ and $\chi_n'=\partial\chi_n\leq \frac{C}{n}$ for some constant $C$. For any $\phi\in\Dd(H^*_0)$ set $\phi_n=\chi_n\phi$. Then $\phi_n\to \phi$ and $H^*_0\phi_n\to H^*_0\phi$ in $L^2(\RR,\CC^{2L})$. Therefore one can calculate as follows:
\begin{eqnarray*}
\langle H^*_0\psi\,|\,\phi\rangle\,-\,\langle \psi\,|\,H^*_0\phi\rangle & = &
\lim_{n,m\to\infty}
\langle H^*_0\psi_n\,|\,\phi_m\rangle\,-\,\langle \psi_n\,|\,H^*_0\phi_m\rangle
\\
& = &
\lim_{n,m\to\infty}
\sum_{j\in \ZZ}\int^{x_j}_{x_{j-1}}dx\;\partial\bigl(\chi_n(x)\chi_m(x)\;
(\Jj\psi(x))^*\phi(x)\,\bigr)
\;,
\end{eqnarray*}
where we used the local integrability of $\Ww$.
This directly implies the proposition.
\hfill $\Box$

\vspace{.2cm}

If $\Pp$ is empty, then the r.h.s. of \eqref{eq-defect} vanishes and this shows that $H_0$ is self-adjoint with domain $W^{1,2}(\RR,\CC^{2L})$. In the terminology of Weyl theory described below, this means that $H$ is in the limit point case for any locally integrable potential $\Ww$. This fact also follows from Weyl theory (more precisely, the bound \eqref{eq-radest} below) without reference to Proposition~\ref{prop-defect}. If $\Pp$ is not empty, then $H^*_0$ has non-trivial deficiency spaces (which are infinite dimensional if and only if $\Pp$ is infinite). Beneath all the self-adjoint extensions of $H_0$ we are interested in those given by local boundary conditions, namely those not mixing the deficiency spaces corresponding to each of the terms on the r.h.s. of \eqref{eq-defect}. Within the class of local boundary conditions we will choose the ones obtained by formally approximating the singular potential $\Vv_j\delta_{x_j}$ (this will be explained below), namely we consider the domain
\begin{equation}
\label{eq-domain}
\Dd(H)\;=\;\left\{\left.
\psi\in W^{1,2}(\RR/\Pp,\CC^{2L})\,\right|\,\psi(x_j+)=e^{\Jj \Vv_j}\psi(x_j-)\,\mbox{ for }j\in\NN\;
\right\}\;.
\end{equation}
Then $H=H_0^*|_{\Dd(H)}$
clearly is an extension of $H_0$ and the identity $(e^{\Jj \Vv_j})^*\Jj e^{\Jj \Vv_j}=\Jj$ replaced in
\eqref{eq-defect} shows that it is self-adjoint.

\vspace{.2cm}

Now that the operator $H$ is well-defined, let us introduce the transfer matrices (or fundamental solutions) $\Tt^z(x,y)\in\,$Mat$(2L\times 2L,\CC)$, $x\geq y\in\RR$, at a complex energy $z\in\CC$ as the unique solutions of
\begin{equation}
\label{eq-transfer}
(H-z)\,\Tt^z(\,.\,,y)\;=\;0\;,\qquad \Tt^z(y,y)\;=\;\one_{2L}\;,
\end{equation}
which are right-continuous in $x$ and in $y$ (for $x\geq y$) and for which
$x\mapsto \Tt^z(x,y)$ is in $\Dd(H)$. (Recall that a function is left-continuous if $f(x-)=f(x)$ for all $x$ and right-continuous if $f(x+)=f(x)$ for all $x$.) For $x<y$, we set $\Tt^z(x,y)=\Tt^z(y,x)^{-1}$. At $x_j\in\Pp$ the transfer matrices then satisfy $\Tt^z(x_j,x_j-)=e^{\Jj \Vv_j}$. The general composition rule reads for $x,u,y\in\RR$
\begin{equation}
\label{eq-transfer2}
\Tt^z(x,y)\;=\;\Tt^z(x,u)\,\Tt^z(u,y)\;.
\end{equation}
For later convenience we also set $\Tt^z(x)=\Tt^z(x,0)$. Now let us briefly sketch in which sense the boundary conditions in \eqref{eq-domain} are natural. Indeed, if $\chi_n\in C_K^\infty(\RR,\RR)$ converges weakly to $\delta_{x_j}$ and $\Tt^z_{n}(x,x')$ is the transfer matrix with potential $\Vv_j\chi_n$, then taking the limit $n\to\infty$ first, one formally verifies
$\Tt^z_{\infty}(x_j,x_j-)=e^{\Jj \Vv_j}$ which is precisely the jump condition above. Next comes the basic but crucial Wronskian identity for the transfer matrices.

\begin{lemma}
\label{lem-Wronski} For $a<b$ and $z,\zeta \in \CC$,
\begin{equation}
\label{eq-Wronski}
\Tt^z(b-)^*\Jj\Tt^\zeta(b-)\;-\;\Tt^z(a)^*\Jj\Tt^\zeta(a)
\;=\; (\zeta-\overline{z})\;\int^b_a dx\;\Tt^z(x)^*\Tt^\zeta(x)\;.
\end{equation}
\end{lemma}

\noindent {\bf Proof.} Denote the points in $\Pp\cup (a,b)$ by $x_1,\ldots,x_N$ and set $x_0=a$ and $x_{N+1}=b$. Then $x\mapsto \Tt^z(x)$ is differentiable away from these points. Thus, using the local integrability of $\Ww$,
\begin{eqnarray*}
(\zeta-\overline{z})\int^b_a \!dx\,\Tt^z(x)^*\Tt^\zeta(x) \!& = &\!
\sum_{j=0}^N\int^{x_{j+1}}_{x_j}\!dx\,\left[\Tt^z(x)^*\bigl((\zeta-\Ww)\Tt^\zeta(x)\bigr)
-\bigl((z-\Ww)\Tt^z(x)\bigl)^*\Tt^\zeta(x)
\right]
\\
\!& = &\!
\sum_{j=0}^N\,\left[\Tt^z(x_{j+1}-)^*\Jj\Tt^\zeta(x_{j+1}-)
\,-\,\Tt^z(x_j+)^*\Jj\Tt^\zeta(x_j+)\right]
\;,
\end{eqnarray*}
where the second equality follows from the differential equation \eqref{eq-transfer} and the fundamental theorem. Replacing $\Tt^\zeta(x_j+)=e^{\Jj \Vv_j}\Tt^\zeta(x_j-)$ and using $(e^{\Jj \Vv_j})^*\Jj e^{\Jj \Vv_j}=\Jj$, one sees that only the boundary terms remain and thus the lemma follows.
\hfill $\Box$

\vspace{.2cm}

Next let us consider the restrictions of $H$ to $\RR_+=(0,\infty)$ and $\RR_-=(-\infty,0)$ given by $H_{\pm}=H|_{L^2(\RR_\pm,\CC^{2L})}$. These operators are not self-adjoint because the same calculation as above shows
\begin{equation}
\label{eq-defect2}
\langle H^*_{\pm}\psi\,|\,\phi\rangle\,-\,\langle \psi\,|\,H^*_{\pm}\phi\rangle\;=\;
\pm\,
\psi(0\pm)^*\Jj\phi(x\pm)
\;,
\end{equation}
for $\psi,\phi\in \Dd(H_{\pm}^*)=
\left\{\left.
\psi\in W^{1,2}(\RR_\pm/\Pp,\CC^{2L})\,\right|\,\psi(x_j+)=e^{\Jj \Vv_j}\psi(x_j-)\,\mbox{ for }j\in\NN\;
\right\}$. This shows that the self-adjoint boundary conditions for $H_\pm$ are precisely given by the set $\LM_L$ of hermitian Lagrangian planes, namely 
$\LM_L=\{\Phi\in\mbox{Mat}(2L\times L,\CC)\,|\,\mbox{rank}(\Phi)=L, \Phi^*\Jj\Phi=0\}/\sim$ where $\Phi\sim\Phi'\;\Leftrightarrow\;\Phi=\Phi'c$ for $c\in\mbox{GL}(L,\CC)$.
For one such plane $\Phi\in\LM_L$, the associated self-adjoint operator will be denoted by $H_{\pm,\Phi}$. It is well-known (see {\it e.g.} \cite{SB1} for a short proof) that $\LM_L$ is diffeomorphic to the unitary group U$(L)$. Thus the deficiency spaces $N^z_{\pm}=\,$ker$(H_{\pm}^*-z)$ of $H_{\pm}$ are $L$-dimensional.

\vspace{.1cm}

For any analytic function $g$ we denote its complex derivative by $\partial_z g=\dot{g}$.

\begin{theo}
\label{theo-WT}
For $\Im m(z)\neq 0$ there exist unique so-called Weyl-Titchmarsh matrices $M^z_\pm\in\,\mbox{\rm Mat}(L\times L,\CC)$ such that $\mbox{\rm ker}(H_{\pm}^*-z)$ is spanned by the column vectors of
\begin{equation}
\label{eq-L2sol}
\Phi_\pm^z(x)\;=\;\Tt^z(x)
\left(\begin{array}{c} \one \\ \pm M^z_\pm\end{array}\right)
\;.
\end{equation}
{\rm (}Here the column vectors of $\Phi_\pm^z$ are considered as elements  of $L^2(\RR_\pm,\CC^{2L})$, but below $\Phi_\pm^z(x)$ is also used for all $x\in\RR$.{\rm )}
They are analytic in $\CC/\RR$ and satisfy the Herglotz property
\begin{equation}
\label{eq-Hergprop}
\frac{\Im m(M^z_\pm)}{\Im m(z)}\,=\,\int_{\RR_\pm}
dx\;\Phi^z_\pm(x)^*\Phi^z_\pm(x)
\,>\,0\;,
\end{equation}
where $\Im m(Z)= \frac{\imath}{2} (Z^*-Z)$ for any operator $Z$,
as well as
$$
(M^z_\pm)^*\,=\,M^{\overline{z}}_\pm\;,\qquad \dot{M}^z_\pm\,=\,
\int_{\RR_\pm}
dx\;\Phi^{\overline{z}}_\pm(x)^*\Phi^z_\pm(x)\;.
$$
\end{theo}

\noindent {\bf Proof.} Let us consider the case of the sign $+$ and $\Im m(z)>0$. It was argued above that the dimension of
$\mbox{\rm ker}(H_{+}^*-z)$ is $L$. As every solution of $H_+\psi=z\psi$ is of the form $\psi(x)=\Tt^z(x)v$ for some vector $v\in\CC^{2L}$, it follows that there are $L\times L$ matrices $\alpha$ and $\beta$ such that the column vectors of
$$
\left(\begin{array}{c} \alpha(x) \\ \beta(x) \end{array}\right)\;=\;\Tt^z(x)
\left(\begin{array}{c} \alpha \\ \beta \end{array}\right)\;,
$$
span $\mbox{\rm ker}(H_{+}^*-z)$.
As these vectors are, in particular, square integrable, replacing them twice in the Wronski identity \eqref{eq-Wronski} with $b=\infty$ and $a=0$ shows that
$$
\imath(\beta^*\alpha-\alpha^*\beta)\;=\;2\, \Im m(z)\;\int_0^\infty dx\;(\alpha(x)^*\alpha(x)+\beta(x)^*\beta(x))
\;>\;0\;.
$$
From this it follows that both $\alpha$ and $\beta$ are invertible because for a vector $v$ in the kernel of $\alpha$ or $\beta$ one would have $v^*(\beta^*\alpha-\alpha^*\beta)v=0$. Therefore one can set $M^z_+=\beta\alpha^{-1}$ and this also leads to the formula \eqref{eq-Hergprop}. The identity $(M^z_\pm)^*=M^{\overline{z}}_\pm$ follows by replacing $\zeta=\overline{z}$ and $a=0$, $b=\infty$ in the Wronski identity \eqref{eq-Wronski}. Finally, let us check the analyticity of $M^z_+$ and derive the formula for its derivative. Again the Wronski identity with $a=0$ and $b=\infty$ shows for $z\neq\zeta$ that
$$
\int^\infty_0 dx\;\Phi_+^{\overline{z}}(x)^*\Phi^\zeta_+(x)
\;=\;
\frac{M^\zeta_+-M^z_+}{\zeta-z}
\;.
$$
Note that the integrand on the l.h.s. is square integrable also in the limit $\zeta\to z$ (at least for $z\in\CC/\RR$), so that $M^z_+$ is indeed holomorphic and the formula for the derivative follows.
The proofs for $M^z_-$ are similar. Let us point out though that due to our definitions the jump $e^{\Jj\Vv_0}$ is relevant for $M^z_-$ if $x_0=0\in\Pp$. This is of some importance below.
\hfill $\Box$

\vspace{.2cm}

As a short aside, let us sketch how the modeling of the singular potential in \eqref{eq-Dirac} by the jump conditions in \eqref{eq-domain} fits with the theory of extensions by von Neumann. For this purpose, let us add the singular potential $\Vv=\Vv_0$ at $x_0=0$ to the operator $H$. Let $\widetilde{H}_0$ be the restriction of $H$ to $\Dd(\widetilde{H}_0)=\{\psi\in\Dd(H)\,|\,\psi(0+)=\psi(0-)=0\}$. Due to Theorem~\ref{theo-WT} the deficiency spaces are both $2L$-dimensional and given by $\mbox{ker}(\widetilde{H}_0-z)={\Psi}^z_+\CC^L\,\oplus\,{\Psi}^z_-\CC^L$, where
$$
{\Psi}^z_\pm(x)\;=\;\chi(\pm x>0)\;
\Tt^z(x) \left(\begin{array}{c} \one \\ \pm M^z_\pm\end{array}\right)
\left( \frac{1}{z-\overline{z}}\,(M^z_\pm-(M^z_\pm)^*)  \right)^{-\frac{1}{2}}\;,
$$
and $\chi$ is the indicator function.
These are partial isometries ${\Psi}^z_\pm:\CC^L\to N^z_\pm$, namely
${\Psi}^z_\pm({\Psi}^z_\pm)^*$ is the projection on $N^z_\pm$ and
$({\Psi}^z_\pm)^*{\Psi}^z_\pm=\one_L$. Now the unitaries from $\mbox{ker}(\widetilde{H}_0-z)$ to $\mbox{ker}(\widetilde{H}_0-\overline{z})$ parameterize the self-adjoint extensions of $\widetilde{H}_0$. Using the partial isometries, these unitaries are precisely given by $({\Psi}^{\overline{z}}_+,{\Psi}^{\overline{z}}_-)U
({\Psi}^{{z}}_+,{\Psi}^{{z}}_-)^*$ where $U$ runs through the unitary group U$(2L)$. It is now a matter of calculation to check that
\begin{equation}
\label{eq-vonNeumannU}
U\;=\;\left[ ({\Psi}^{\overline{z}}_+(0+),0)- e^{\Jj \Vv}(0,{\Psi}^{\overline{z}}_-(0-))\right]^{-1}
\left[ ({\Psi}^{{z}}_+(0+),0)- e^{\Jj \Vv}(0,{\Psi}^{{z}}_-(0-))\right]
\,,
\end{equation}
is well-defined ({\it i.e.} the inverse exists), is unitary and gives exactly the self-adjoint extension given by the jump condition $\psi(0+)=e^{\Jj \Vv}\psi(0-)$. Hence every local boundary condition in \eqref{eq-domain} is an extension within the local $2L$-dimensional deficiency spaces in the sense of von Neumann. On the other hand, there are local von Neumann extensions which are not given by jump conditions (for example, those which do not couple left and right).

\vspace{.2cm}

Even though it was already shown above that $H$ is always self-adjoint
(so that one is always in the limit point case), we now describe the
Weyl theory because it gives quantitative estimates for the
Weyl-Titchmarsh matrices needed below. We closely stick to the
notations of our prior work \cite{SB2} along the lines of which also
the proofs of the results below can be given (even though there are
definitely older references such as \cite{HS} for some of them). The
basic idea is to study the restriction of the operator $H_+$ to
${L^2((0,x),\CC^{2L})}$ and to analyze which initial conditions at $0$
lead to solutions satisfying any self-adjoint boundary conditions at
$x$ (there is an analogous treatment for $H_-$). If an adequate chart
for these initial conditions is used they have the geometric structure
of a matrix circle in the upper half-plane, called the Weyl
surface. As $x$ increases, this circle shrinks in a nested manner. In
the so-called limit point case that one always encounters for the
Dirac operators, it shrinks to a single point in the limit
$x\to\infty$ identified with the initial condition of \eqref{eq-L2sol}
specified by the Weyl-Titchmarsh matrix $M_+^z$. This fact reflects
that there is no need to fix a boundary conditions at infinity in this
case (the $L^2$-condition takes care of it) because $H$ is already
self-adjoint. 

\vspace{.2cm}

Now comes the more technical description.  Let $\GM_L$ be the Grassmannian of $L$-dimensional planes in $\CC^{2L}$. The chart on $\GM_L$ used is the stereographic projection $\pi$  sending
an $2L\times L$ matrix $\binom{\alpha}{\beta}$ representing the plane to $\alpha\beta^{-1}\in\,$Mat$(L,\CC)$. It is defined on full measure subset $\GM_L^\inv\subset\GM_L$
on which the inverse of $\beta$ exists. Then the Weyl surface at $x\neq 0$ is defined by
$$
\partial \WM^z(x)\;=\;-\,\pi\left(\left\{
\Phi\in\GM_L\,\left|\,\Tt^z(x)\,\Phi\in\LM_L\,
\right.\right\}\right)
\;=\;\left\{-M^{-1}\,\left|\,\Tt^z(x)\,\left(\begin{array}{c} \one \\ M\end{array}\right)\in\LM_L\,
\right.\right\}\;,
$$
where the equality follows by showing that every plane $\Phi$ in the first set is of the form in the second one \cite[Prop.~7]{SB2}. Now it is useful to rewrite the condition $\Tt^z(x)\,\Phi\in\LM_L$ in terms of the quadratic form
$$
\Qq^z(x)\;=\;\frac{1}{\imath}\;\Tt^z(x)^*\Jj\Tt^z(x)\;,
$$
namely $\partial \WM^z(x)=-\,\pi\left(\left\{
\Phi\in\GM_L\,\left|\,\Phi\mbox{ isotropic for }\Qq^z(x)\,
\right.\right\}\right)$. The definition of $\Qq^z(x)$ shows that $\Qq^z(x+)=\Qq^z(x-)$ also for $x\in\Pp$ so that $\Qq^z(x)$ is continuous and thus $\partial \WM^z(x+)=\partial \WM^z(x-)$. Item (i) and (ii) of the following properties of $\Qq^z(x)$ follow from the definition and the Wronskian identity, while (iii) can be checked as in \cite{SB2} once one has verified that $\Tt^z(x)^{-1}=\Jj^*\Tt^{\overline{z}}(x)\Jj$.

\begin{proposi}
\label{prop-Qproperties} The quadratic form $\Qq^z(x)$ satisfies:

\vspace{.1cm}

\noindent {\rm (i)} $\Qq^z(x)=\frac{1}{\imath}\,\Jj+2\,\Im m(z)\,
\langle\Tt^z(\,.\,)| \Tt^z(\,.\,)\rangle_{L^2((0,x),\CC^{2L})}$

\vspace{.1cm}

\noindent {\rm (ii)} $\Im m(z)\,\partial\Qq^z(x)\geq 0$

\vspace{.1cm}

\noindent {\rm (iii)} $\Qq^z(x)^{-1}=\Jj^*\Qq^{\overline{z}}(x)\Jj$

\end{proposi}

Now the radial and center operator are defined by
$$
R^z(x)\;=\;\left[\,\left(\begin{array}{c} \one \\ 0\end{array}\right)^*\,\Qq^z(x)\,
\left(\begin{array}{c} \one \\ 0\end{array}\right)\,\right]^{-1}\;,\qquad
S^z(x)\;=\;R^z(x)\,\left(\begin{array}{c} \one \\ 0\end{array}\right)^*\,\Qq^z(x)\,
\left(\begin{array}{c} 0 \\ \one \end{array}\right)\;.
$$
Both $R^z(x)$ and $S^z(x)$ are continuous in $x$ (apart from the singularity at $x=0$).
It follows from item (i) of Proposition~\ref{prop-Qproperties} that $R^z(x)>0$ and $-R^{\overline{z}}(x)>0$ for $\Im m(z)>0$, and item (ii) implies $\partial R^z(x)\leq 0$. The terms radial and center operator are justified by the following result which can be checked by the same calculation as in \cite{SB2}. It is the basic fact of Weyl theory. Let the matrix upper half-plane $\UM_L$ be defined as the set of matrices $Z\in\,$Mat$(L,\CC)$ satisfying $\Im m(Z)>0$.

\begin{theo}
\label{theo-Weyl}
Let $\Im m(z)> 0$. Then
$$
\partial\WM^z(x)\;=\;\left\{\left.\,S^z(x)+R^z(x)^{\frac{1}{2}}U
(-R^{\overline{z}}(x))^{\frac{1}{2}}\;\right|\;U^*U=\one\,\right\}\;\subset\;\UM_L\;.
$$
If now the open and closed Weyl disc $\WM^z(x)$ and $\overline{\WM^z(x)}$ are defined by this formula with $U$ running through the set defined by $U^*U<\one$ and $U^*U\leq \one$ instead of the unitary group $\mbox{\rm U}(L)$, then the Weyl surfaces are strictly nested in the sense that for $x>x'>0$ or $x<x'<0$
$$
\WM^z(x)\,\subset\,\WM^z(x')\;,\qquad \partial\WM^z(x')\,\cap\,\overline{\WM^z(x)}\;=\;\emptyset\;.
$$
\end{theo}

\vspace{.2cm}

This theorem can also be used to prove the uniqueness of $M^z_+$ instead of the above argument based on \eqref{eq-defect2}, that is, basically the calculation in the proof of Proposition~\ref{prop-defect}. Indeed, along the lines of Proposition~11 of \cite{SB2} one can prove that there exists a constant $c$ such that
\begin{equation}
\label{eq-radest}
\|R^z(x)\|\;\leq\;\frac{c}{|x|\;\Im m(z)^2}\;.
\end{equation}
This implies that $H_\pm$ is in the limit point case in the literal sense and that one furthermore has $-(M^z_\pm)^{-1}=\lim_{x\to\pm\infty}S^z(x)$. We need the following consequence for our purposes below. It replaces perturbative arguments in \cite{KS,Sun} and hence the bounds below hold under the more natural assumptions that $\Ww$ is locally integrable. For Schr\"odinger operators a similar reasoning applies if they are supposed to be in the limit point case.

\begin{coro}
\label{coro-Mbounds} There are constants $c_1,c_2$ depending only on $z$ and the $L^1_\loc$-norm of $\Ww$ such that
$$
\|M^z_\pm\|\;\leq\;c_1\;,\qquad \frac{1}{c_2}\;\leq\;
\frac{\Im m(M^z_\pm)}{\Im m(z)}\;\leq \;{c_2}\;.
$$
\end{coro}

\noindent {\bf Proof.} At $x=0$ the radial operator is infinite in the sense that $R^z(0)^{-1}=0$. As
$$
\partial (R^z(x)^{-1})\;=\;\Im m(z)\;
\left(\begin{array}{c} \one \\ 0 \end{array}\right)^*\Tt^z(x)^*\Tt^z(x)
\left(\begin{array}{c} \one \\ 0 \end{array}\right)
$$
is equal to $\Im m(z)\one>0$ for $x=0$ and is continuous in $x$ (even differentiable), it follows that
$R^z(x)^{-1}>0$ for some $x>0$. Hence $\|R^z(x)\|<\infty$ and the Weyl disc $\overline{\WM^z(x)}$ is compact and strictly contained in the upper half-plane $\UM_L$. Furthermore by Theorem~\ref{theo-Weyl} the limit point $-(M^z_\pm)^{-1}$ is an element of $\overline{\WM^z(x)}$. As $Z\mapsto -Z^{-1}$ maps compact sets of $\UM_L$ to compact sets of $\UM_L$ the proof is complete.
\hfill $\Box$

\section{Green's function and spectral analysis}
\label{sec-Green}

This section deals with the Green function and spectral theory of the
self-adjoint operator \eqref{eq-Dirac} defined by
\eqref{eq-domain}. We always assume that $x_0=0\in\Pp$, set
$\Vv=\Vv_0$ and denote the operator with singular potential $\Vv$ by
$H_\Vv$ (hoping that the reader can distinct $H_0$ with $\Vv=0$ from
the $H_0$ in the last section).

\begin{proposi}
\label{prop-Greenwithout}
Let $\Im m(z)\neq 0$ and $M_\pm^z$, $\Tt^z(x)$ and $\Phi^z_\pm$ be associated to $H_0$ {\rm (}this only leads to changes for $x<0$ and the sign $-${\rm )}. The resolvent $(H_0-z)^{-1}$ is an integral operator with kernel
\begin{equation}
\label{eq-Greenwithout}
G_0^z(x,y)
\;=\; \Phi^z_\pm(x) \,
(-M^z_+-M^z_-)^{-1}\,\Phi_\mp^{\overline{z}}(y)^*
\;,
\end{equation}
where the upper and lower signs are taken if $x<y$ and $x>y$ respectively. Furthermore, for a Lagrangian plane $\Phi=(\one\;\gamma)^*$, the resolvent $(H_{+,\Phi}-z)^{-1}$ is an integral operator with kernel
$$
G_{+,\Phi}^z(x,y)
\;=\;
\left\{\begin{array}{cc}
\Tt^z(x)\,\Phi\,(-M^z_++\gamma)^{-1}\,\Phi_+^{\overline{z}}(y)^*
\;, & x<y\;, \\
& \\
\Phi^z_+(x) \,
(-M^z_++\gamma)^{-1}\,\Phi^*\,\Tt^{\overline{z}}(y)^*
\;, & x>y\;.
\end{array}\right.
$$
\end{proposi}

\noindent {\bf Proof.} Let $G_0^z$ be defined by the formula in the theorem. Using
$(M^{\overline{z}}_\mp)^*=M^{{z}}_\mp$ one readily verifies that for all $x\in\RR$,
\begin{equation}
\label{eq-Greenjump}
\lim_{\epsilon\downarrow 0} \bigl[\,G^z_0(x+\epsilon,x)-G_0^z(x-\epsilon,x)\, \bigr]\;=\; \Tt^z(x) \Jj\Tt^{\overline{z}}(x)^* \;=\;\Jj\;,
\end{equation}
where the last equality follows by taking the inverse of  $\Tt^{\overline{z}}({x})^* \Jj\Tt^{{z}}(x)=\Jj$, which is the Wronskian identity \eqref{eq-Wronski} with $\zeta=\overline{z}$, $a=0$ and $b=x$. Therefore setting $\psi(x)=\int dy\, G^z_0(x,y)\phi(y)$ for a smooth function $\phi\in L^2(\RR,\CC^{2L})$, the definition \eqref{eq-transfer} of the transfer matrices implies that $(H_0-z)\psi=\phi$ because $\partial\,$sgn$=2\delta_0$ if sgn is the sign function and $\delta_x$ is a Dirac peak at $x$. Hence $G^z_0$ is indeed the desired integral kernel. The formula for the half-sided operator is verified in a similar manner.
\hfill $\Box$

\vspace{.2cm}

From Proposition~\ref{prop-Greenwithout}, \eqref{eq-vonNeumannU} and the general Krein formula for resolvents of self-adjoint extensions one could now deduce an explicit formula for the integral kernel $G_\Vv(x,y)$ of $H_\Vv$. Then lengthy algebraic calculations lead to Proposition~\ref{prop-Greenwith} below, but we can also deduce it more directly based on the following idea. Both functions $x\mapsto G_\Vv^z(x,y)$ and $y\mapsto G_\Vv^z(x,y)^*=G_\Vv^{\overline{z}}(y,x)$ are in the domain $\Dd(H_\Vv)$ and satisfy respectively
$(H_\Vv-z)G_\Vv^z(.,y)=\delta_y$ and $(H_\Vv-\overline{z})G_\Vv^z(x,.)=\delta_x$. Away from $x_0=0$, the domain of $\Dd(H_0)$ and the identities for $H_0$ are the same. Thus a good Ansatz is
$$
G^z_\Vv(x,y)\;=\;G^z_0(x,y)\,+\,G^z_0(x,0+)\,\Kk\,G^z_0(0-,y)\;,
$$
with a matrix $\Kk$ to be determined. The jump condition $G^z_\Vv(0+,y)=e^{\Jj \Vv}G^z_\Vv(0-,y)$ gives for $y\neq 0$
$$
G^z_0(0,y)\,+\,G_0^z(0+,0)\,\Kk G_0^z(0,y)\;=\;
e^{\Jj \Vv}\left[G^z_0(0,y)\,+\,G_0^z(0-,0)\,\Kk G_0^z(0,y)\right]\;.
$$
Now let us take the difference of this equation for $y=0+$ and $y=0-$. Because $G^z_0(0+,0)-G^z(0-,0)=\Jj$ by \eqref{eq-Greenjump}, one obtains
$$
\Jj\,+\,G^z_0(0+,0)\,\Kk\Jj\;=\;e^{\Jj \Vv}\left[\Jj\,+\,G^z_0(0-,0)\,\Kk\Jj\right]
$$
This equation can formally be solved for $\Kk$, leading to the following formula.

\begin{proposi}
\label{prop-Greenwith}
Let $\Im m(z)\neq 0$. The resolvent $(H_\Vv-z)^{-1}$ is an integral operator with kernel
\begin{equation}
\label{eq-resolventid}
G_\Vv^z(x,y)
\;=\;
G_0^z(x,y)
\,+\,
G_0^z(x,0)
\left[
e^{\Jj \Vv}G_0^z(0-,0)\,-\,
G_0^z(0+,0)
\right]^{-1}\,(1-e^{\Jj \Vv})\,
G_0^z(0,y)\;.
\end{equation}
\end{proposi}

\noindent {\bf Proof.} It remains to check that the appearing inverse is indeed well-defined.
Due to \eqref{eq-Greenwithout}, there exist two $L$-dimensional planes
$\Phi_\pm$ with $\pm\pi(\Phi_\pm) \in\UM_L$ such that
$G_0^z(0-,0)=\Phi_+\Phi_-^*$ and $G_0^z(0+,0)=\Phi_-\Phi_+^*$. Now we
claim that for any hermitian symplectic $\Tt$ satisfying
$\Tt^*\Jj\Tt=\Jj$, thus in particular $\Tt=e^{\Jj \Vv}$, one has $\Tt\Phi_+\CC^L\cap\Phi_-\CC^L=\{0\}$.
This implies as desired that $\Tt\Phi_+\Phi_-^*-\Phi_-\Phi_+^*$ is
invertible. To prove the claim we first note that
$\pi(\Tt\Phi_+)\in\UM_L$ (as the M\"obius transformation with a
hermitian symplectic matrix
sends $\UM_L$ to $\UM_L$) so that it is sufficient to
consider the case $\Tt=\one$. Now let $\Phi_+v=\Phi_-w$ for some
$v,w\in\CC^L$. Set $\alpha_\pm=(\one\;0)\Phi_\pm$ and
$\beta_\pm=(0\;\one)\Phi_\pm$, both of which are known to be
invertible. Then $\alpha_+v=\alpha_-w$ and $\beta_+v=\beta_-w$. Thus
$v=\beta_+^{-1}\beta_-w$ so that
$\alpha\beta_+^{-1}\beta_-w=\alpha_-w$. Therefore $u=\beta_-w$
satisfies $\alpha_+\beta_+^{-1}u = \alpha_-\beta_-^{-1}u$ and thus
$u^*\pi(\Phi_+)u=u^*\pi(\Phi_-)u$. By hypothesis this implies $u=0$
and consequently $w=v=0$. 
\hfill $\Box$

\vspace{.2cm}

Before going on let us discuss the discontinuities of $G_\Vv^z(x,y)$
in the vicinity of the point $(x,y)=(0,0)$ (any other singular point
can be analyzed similarly). Because $x\mapsto G_\Vv^z(x,y)$ and
$y\mapsto G_\Vv^z(x,y)^*=G_\Vv^{\overline{z}}(y,x)$ are in the domain
$\Dd(H_\Vv)$, the singular potential leads to jumps on the lines $x=0$
and $y=0$. According to \eqref{eq-Greenjump} there is furthermore a
jump by $\Jj$ on the diagonal $x=y$. Away from these 3 lines crossing
at the origin, $G_\Vv^z(x,y)$ is continuous. Hence there are 6
directional limits as $(x,y)\to (0,0)$. Enumerate them by
$G_1,\ldots,G_6$ in a clockwise direction starting with
$G_1=\lim_{\epsilon\downarrow 0} G_\Vv^z(\epsilon,2\epsilon)$. Setting
$\Tt=e^{\Jj\Vv}$ one then has 
$$
G_2=G_1+\Jj\,,\;\;\;G_3=G_2(\Tt^{-1})^*\,,\;\;\;
G_4=\Tt^{-1} G_3\,,\;\;\;G_5=G_4-\Jj\,,\;\;\;
G_6=G_5\Tt^*\,,\;\;\;G_1=\Tt G_6\,.
$$
Note that these relations are indeed cyclic because
$\Tt^*\Jj\Tt+\Jj$. By \eqref{eq-Greenwithout} each
of the $G_j$ has rank $L$. The following proposition shows that,
however, an adequate linear combination is a Herglotz function and, in
particular, of full rank $2L$. 

\begin{proposi}
\label{prop-GreenHerg}
Let us define the averaged Green matrix
\begin{eqnarray*}
\widehat{G}_\Vv^z(x)
& = & \lim_{\epsilon\downarrow 0}\left[\;
\frac{1}{4}\,G_\Vv^z(x+\epsilon,x-\epsilon)
\,+\,\frac{1}{4}\,
G_\Vv^z(x-\epsilon,x+\epsilon)
\,+\,\frac{1}{8}\,
G_\Vv^z(x+\epsilon,x+2\epsilon)\right.
\\
& & \;\;\;\;\;\; \left. +\,\frac{1}{8}\,
G_\Vv^z(x+2\epsilon,x+\epsilon)
\,+\,\frac{1}{8}\,
G_\Vv^z(x-\epsilon,x-2\epsilon)
\,+\,\frac{1}{8}\,
G_\Vv^z(x-2\epsilon,x-\epsilon)\,
\right]\;.
\end{eqnarray*}
Then
$z\in\UM_1\mapsto\widehat{G}_\Vv^z(x)=(\widehat{G}_\Vv^{\overline{z}}(x))^*
\in\mbox{\rm Mat}(2L,\CC)$ is a Herglotz function for any $x\in\RR/\Pp$ and has non-negative imaginary part for $x\in\Pp$. It satisfies
\begin{equation}
\label{eq-GreenIm}
\Im m(\widehat{G}_\Vv^z(0))\;=\;\frac{1}{4}\;(\one+e^{\Jj \Vv})\,\Im m(\widehat{G}_\Vv^z(0+))\,(\one+e^{\Jj \Vv})^*\;.
\end{equation}
\end{proposi}

\noindent {\bf Proof.} Let us note that for $x\notin\Pp$ the definition of the averaged Green matrix reduces to $\widehat{G}_\Vv^z(x)=\frac{1}{2}(G^z_\Vv(x+,x)+G^z_\Vv(x-,x))$.
For sake of notational simplicity, let us focus on the case $x=0$ with $\Vv\not = 0$ modeling $x\in\Pp$. With the above notations, then by definition $\widehat{G}_\Vv^z(0)=\frac{1}{8}(G_1+G_2+2G_3+G_4+G_5+2G_6)$ which is a weighing of the $G_j$ according to the area of the corresponding octant or quadrant. Now let $z=E+\imath\epsilon$ with $\epsilon>0$ and consider the positive operator $\Im m\bigl((H_\Vv-z)^{-1}\bigr)=\epsilon ((H_\Vv-E)^2+\epsilon^2)^{-1}$. For any $\varphi\in L^2(\RR,\CC^{2L})$, one thus has
$$
0\;<\;\langle\varphi|\Im m\bigl((H_\Vv-z)^{-1}\bigr)|\varphi\rangle
\;=\;\frac{1}{2\imath}\;\int dx\int dy\;\varphi(x)^*\bigl(
G^z_\Vv(x,y)-G^z_\Vv(y,x)^*\bigr)\varphi(y)\;.
$$
Now let $\chi_k\in C_K^\infty(\RR)$ be a positive approximate unit, that is $w$-$\lim_{k\to\infty}\chi_k=\delta_0$. For any function $f:\RR^2\cong\CC\to\CC$ having the
directional limits $f(\theta)=\lim_{r\downarrow 0} f(re^{\imath\theta})$, it follows that $\int dx\int dy\,\chi_k(x)\chi_k(y) \,f(x,y)$ converges to $\hat{f}=\int^{2\pi}_0 \frac{d\theta}{2\pi}\,f(\theta)$. Hence, for $\varphi_k=\chi_kv$ with $v\in\CC^{2L}$,
$$
0\;\leq\;\lim_{k\to\infty}\,
\langle\varphi_k|\Im m\bigl((H_\Vv-z)^{-1}\bigr)|\varphi_k\rangle
\;=\;
\frac{1}{2\imath}\;v^*\left(\widehat{G}^z_\Vv(0)-\widehat{G}^z_\Vv(0)^*\right)v\;.
$$
This proves that the imaginary part is non-negative. The Herglotz property for $0\notin\Pp$, namely that the imaginary part is positive, follows from the concrete formula
\begin{equation}
\label{eq-G0av}
\widehat{G}^z_0(0)\;=\;
\left(\begin{array}{cc}
(-M_+-M_-)^{-1} &  (-M_+-M_-)^{-1}(M_+-M_-) \\
(M_+-M_-)(-M_+-M_-)^{-1} & (M_+^{-1}+M^{-1})^{-1}
\end{array}\right)
\end{equation}
following from Proposition~\ref{prop-Greenwithout}, and the Herglotz property of $M_\pm$ by the Liouville theorem.
As the singular points are discrete, there is an interval
$(0,\epsilon)$ not containing any. Hence
$\widehat{G}_\Vv^z(0+)=\frac{1}{2}(G_1+G_2)$. It is now a matter of an
algebraic calculation to verify the second formula. 
\hfill $\Box$

\vspace{.2cm}

As for any Herglotz function with sufficient decay properties such as $\widehat{G}_\Vv^z(x)$, there is associated a matrix valued measure $\mu_x$ on $\RR$ and a self-adjoint matrix $A_x=A_x^*$ independent of $z$ (see \cite{GT} for a review and properties) such that
$$
\widehat{G}_\Vv^z(x)\;=\;A_x\,+\,\int \mu_x(dE)\;\left(\frac{1}{E-z}\,-\,\frac{1}{1+E^2}\right)\;.
$$
Because
$$
\widehat{G}^z_\Vv(x)\;=\;\Tt^z(x,y)\,\widehat{G}^z_\Vv(y)\,\Tt^z(x,y)^*
$$
for $x,y\notin\Pp$ and $\Tt^z(x,y)$ is analytic and invertible, the measures $\mu_x$, $x\notin\Pp$, all define the same measure class. According to \eqref{eq-GreenIm}, the measure $\mu_0$ associated to
$\widehat{G}^z_\Vv(0)$ is also in the same measure class as long as $-1$ is not in the spectrum of $e^{\Jj \Vv}$. We skip the proof of the following result, showing in which sense $\mu_x$ can rightfully be called a spectral measure of $H_\Vv$ (see \cite{KS}).

\begin{proposi}
\label{prop-specmeas}
Let $\psi,\phi\in L^2(\RR,\CC^{2L})$ and $f\in C_0(\RR)$. Then, whenever $\mu_x$ is in the almost sure measure class,
$$
\langle \psi\,|\,f(H_\Vv)\,|\,\phi\rangle
\;=\;
\int_\RR\,f(E)\;\left(\int dy\,\Tt^E(y,x)^*\psi(y)\,\right)^*\,\mu_x(dE)\,
\left(\int dy\,\Tt^E(y,x)^*\phi(y)\,\right)
\;,
$$
and the functions of $E$ in the parenthesis are in $L^2(\RR,\mu_x)$.
\end{proposi}

The arguments in Section~\ref{sec-Absence} will be based on the following perturbative formula for the averaged Green matrix w.r.t. the finite rank perturbation given by the singular potential $\Vv\delta_0$. For notational convenience let us set $\widehat{G}_\Vv^z=\widehat{G}_\Vv^z(0)$. Furthermore let us introduce the Cayley transform of $\Vv$ by
\begin{equation}
\label{eq-Vhat}
\widehat{\Vv}\;=\;2\,\Jj(e^{\Jj \Vv}+\one)^{-1}(e^{\Jj \Vv}-\one)\;,
\end{equation}
whenever the inverse is well-defined. One readily checks that $\widehat{\Vv}^*=\widehat{\Vv}$ and that $\Jj^*\widehat{\Vv}^t\Jj=\widehat{\Vv}$ if $\Jj^* \Vv^t\Jj=\Vv$.

\begin{proposi}
\label{prop-GreenPerturb}
The averaged Green matrix satisfies {\rm (}even if $\widehat{\Vv}$ is not well-defined{\rm )}
\begin{equation}
\label{eq-GreenPerturb1}
\widehat{G}_\Vv^z
\; =\;
\left[\,(\widehat{G}_0^z)^{-1}\,+\,\widehat{\Vv}\,\right]^{-1}\;,
\end{equation}
and
\begin{equation}
\label{eq-GreenPerturb2}
\Im m(\widehat{G}_\Vv^z)
\; =\;
\left(\left[\,\one\,+\,\widehat{\Vv}\,\widehat{G}_0^z\,\right]^{-1}\right)^*
\Im m(\widehat{G}_0^z)
\left[\,\one\,+\,\widehat{\Vv}\,\widehat{G}_0^z\,\right]^{-1}
\;.
\end{equation}
\end{proposi}

\noindent {\bf Proof.}
Let us apply the averaging procedure of Proposition~\ref{prop-GreenHerg} to \eqref{eq-resolventid}. This gives
$$
\widehat{G}^z_\Vv\;=\;\widehat{G}^z_0\,+\,\widehat{G}^z_0\,\Kk\,\widehat{G}^z_0\;=\;\widehat{G}^z_0\,
\bigl(\one\,+\,\Kk\,\widehat{G}^z_0\bigr)\;,
$$
where $\Kk=\left[e^{\Jj \Vv}G_0^z(0-,0)-G_0^z(0+,0) \right]^{-1}(\one-e^{\Jj\Vv})$ as before.
Because both $\widehat{G}^z_\Vv$ and $\widehat{G}^z_0$ are invertible, it follows that also
$\bigl(\one\,+\,\Kk\,\widehat{G}^z_0\bigr)$ is invertible. Hence
\begin{equation}
\label{eq-GreenPerturb3}
\widehat{G}^z_\Vv\;=\;\widehat{G}^z_0\,+\,\widehat{G}^z_\Vv\,\bigl(\one\,+\,\Kk\,\widehat{G}^z_0\bigr)^{-1}\,
\Kk\,\widehat{G}^z_0\;=\;
\left[
(\widehat{G}^z_0)^{-1}-\bigl(\one\,+\,\Kk\,\widehat{G}^z_0\bigr)^{-1}\,
\Kk\,\right]^{-1}\;.
\end{equation}
Using $G_0^z(0\pm,0)=\widehat{G}^z_0\pm \frac{1}{2}\,\Jj$, one readily checks
$\bigl(\one +\Kk\,\widehat{G}^z_0\bigr)^{-1}\Kk=-\widehat{\Vv}$ completing the proof of
\eqref{eq-GreenPerturb1}. That of \eqref{eq-GreenPerturb2} is straightforward.
\hfill $\Box$

\section{Stochastic Dirac operators}
\label{sec-random}

In this section we introduce stochastic Dirac operators and state a few of their elementary properties, then introduce the random Dirac operators and give a precise statement of the main coupling hypothesis needed in Theorem~\ref{theo-main}. Let be given a compact dynamical system $(\Omega,\PP,T)$ where $T$ is a continuous $\RR$-action on the compact space $\Omega$ w.r.t. which the probability measure $\PP$ is supposed to be ergodic. Then $(H(\omega))_{\omega\in\Omega}$ is called a family of stochastic Dirac operators if each $H(\omega)$ is of the form \eqref{eq-Dirac} and the map $\omega\in\Omega\mapsto H(\omega)$ is strongly continuous in the resolvent sense and covariant, that is, if $U_x$ denotes the right shift by $x$ on $L^2(\RR,\CC^{2L})$, then $U_x (H(\omega)-z)^{-1}U_x^*=(H(T_x\omega)-z)^{-1}$. Each point $\omega\in\Omega$ is thought of as a configuration, incorporating the positions $\Pp$ and values $(\Vv_x)_{x\in\Pp}$ of the singular potential as well as the potential $\Ww$. Thus $\Pp$ is an $\RR$-ergodic point process. Its density is denoted by $\rho_\Pp$.
The locally integrable potential associated to a given configuration
$\omega$ is then $\Ww_\omega(x)=\Ww(T_{-x}\omega)$, $x\in\RR$, where
the $\Ww$ is a matrix-valued function on $\Omega$. Hence we suppose
this function $\Ww$ to be locally integrable along orbits with a
uniform bound on the $L^1$-norm over unit intervals.

\vspace{.2cm}

Now all objects such as transfer matrices, Weyl-Titchmarsh matrices and Green matrices analyzed in the sections above depend on $\omega$; however, in the notations this will be made explicit by a supplementary argument only if necessary. Let us introduce some notations for the $L\times L$ matrix entries of the potential:
$$
\Ww\;=\;\left(\begin{array}{cc} P & R \\ R^* & Q \end{array}\right)
\;,\qquad
e^{\Jj \Vv}\;=\;\left(\begin{array}{cc} A & B \\ C & D \end{array}\right)
\;.
$$
All these objects are random and for $\Vv=\Vv_x$, $x\in\Pp$, the entries are also denoted $A_x,B_x,C_x,D_x$.
As the matrix $e^{\Jj \Vv}$ is in SP$(2L,\CC)$, it is well-known that the inverse in the definition of the M\"obius transformation
$$
\left(\begin{array}{cc} A & B \\ C & D \end{array}\right)\cdot Z\;=\;(AZ+B)(CZ+D)^{-1}\;,
$$
exists whenever $Z$ is in the upper or lower half-plane, {\it i.e.} $\pm\Im m(Z)>0$. If then $W=e^{\Jj\Vv}\cdot Z=(AZ+B)(CZ+D)^{-1}$, also $W$ is in the upper or lower half-plane respectively and one has $Z=(e^{\Jj\Vv})^{-1}\cdot W=(D^*-B^*W)(-C^*+A^*Z)^{-1}$.
Now we can collect a few first properties of the transfer matrices and the Weyl-Titchmarsh matrices.

\begin{lemma}
\label{lem-Ricatti}
Let $\Im m(z)\neq 0$, set
\begin{equation}
\label{eq-alphabeta}
\left(\begin{array}{c}\alpha^z_\pm(x,\omega) \\ \beta^z_\pm(x,\omega)\end{array}\right)
\;=\;\Tt^z(x,\omega)\,
\left(\begin{array}{c}\one \\ \pm M^z_\pm(\omega)\end{array}\right)
\;=\;\Phi^z_\pm(x,\omega)
\;.
\end{equation}

\vspace{.1cm}

\noindent {\rm (i)}
The transfer matrices satisfy the cocycle equation
$$
\Tt^z(x+y,\omega)\;=\;\Tt^z(x,T_{-y}\omega)\,\Tt^z(y,\omega)\;,\qquad \Tt^z(0,\omega)\;=\;\one\;.
$$

\vspace{.1cm}

\noindent {\rm (ii)} One has
$$
\left(\begin{array}{c}\alpha^z_\pm(x+y,\omega) \\ \beta^z_\pm(x+y,\omega)\end{array}\right)
\;=\;
\left(\begin{array}{c}\alpha^z_\pm(x,T_{-y}\omega) \\ \beta^z_\pm(x,T_{-y}\omega)\end{array}\right)
\;\alpha^z_\pm(y,\omega)\;.
$$
\indent In particular, $\alpha^z_\pm(x,\omega)$ is a cocycle:
$$
\alpha^z_\pm(x+y,\omega)\;=\;
\alpha^z_\pm(x,T_{-y}\omega)\,\alpha^z_\pm(y,\omega)\;,
\qquad \alpha_\pm^z(0,\omega)\;=\;\one\;.
$$

\vspace{.1cm}

\noindent {\rm (iii)}
$
M^z_\pm(T_{-x}\omega)\;=\;\pm\,\beta^z_\pm(x,\omega)\,\alpha^z_\pm(x,\omega)^{-1}\;.
$

\vspace{.1cm}

\noindent {\rm (iv)} The map $x\mapsto M^z_\pm(T_x\omega)$ is differentiable away from $\Pp$. It is left-continuous and for $-x\in\Pp$,
$$
\pm M^z_\pm(T_{x+}\omega)^{-1}
\;=\;(e^{\Jj\Vv_{-x}})^{-1}\cdot (\pm M^z_\pm(T_x\omega)^{-1})
\;.
$$

\vspace{.1cm}

\noindent {\rm (v)} The maps $y\mapsto \alpha^z_\pm(x,T_y\omega)$ and $y\mapsto \beta^z_\pm(x,T_y\omega)$ are left-continuous. For $-y\in\Pp$,
$$
\left(\begin{array}{c} \alpha^z_\pm(x,T_{y+}\omega) \\
\beta^z_\pm(x,T_{y+}\omega) \end{array}\right)
\;=\;(e^{\Jj\Vv_x})^{-1}\,
\left(\begin{array}{c} \alpha^z_\pm(x,T_y\omega) \\
\beta^z_\pm(x,T_y\omega) \end{array}\right)
\,(D_{-y}^*-B_{-y}^*(\pm M^z_\pm(T_y\omega)))^{-1}
\;.
$$

\vspace{.1cm}

\noindent {\rm (vi)} The map $x\in\RR_+\mapsto \alpha^z_\pm(x,\omega)$ is right-continuous. If $x\in\Pp$,
$$
\alpha_\pm^z(x,\omega)\;=\;
(A_x\pm B_xM^z_\pm(T_{-x+}\omega))\,\alpha^z_\pm(x-,\omega)\;=\;
(D^*_x\mp B_x^*M^z_\pm(T_{-x}\omega)\bigr)^{-1}\,\alpha^z_\pm(x-,\omega)\;.
$$

\vspace{.1cm}

\noindent {\rm (vii)} $\partial_x\alpha^z_\pm(x,\omega)=\bigl[-R(T_{-x}\omega)^*\mp(Q(T_{-x}\omega) -z)M^z_\pm(T_{-x}\omega)\bigr]\alpha^z_\pm(x,\omega)$ for $x\notin\Pp$

\vspace{.1cm}

\noindent {\rm (viii)} The following Ricatti equation holds for $x\notin \Pp$
$$
\pm\,\partial_x M^z_\pm(T_{-x}\omega)\;=\;
\left(\begin{array}{c} \one \\\pm M^{\overline{z}}_\pm(T_{-x}\omega) \end{array}\right)^*(\Ww(T_{-x}\omega)-z)
\left(\begin{array}{c} \one \\ \pm M_\pm^z(T_{-x}\omega) \end{array}\right)\;.
$$

\end{lemma}

\noindent {\bf Proof.} (i), (ii) and (iii) follow immediately from \eqref{eq-transfer2} and \eqref{eq-alphabeta}.
It is clearly sufficient to analyze the directional continuity in (iv) and (v) for the case $x=0\in\Pp$.
Let $\epsilon>0$. Using the composition rule for transfer matrices and their translation property
$$
\Tt^z(x+\epsilon,y+\epsilon,T_\epsilon\omega)\;=\;\Tt^z(x,y,\omega)\;,
$$
one deduces
$$
\Tt^z(x,\omega)\;=\;\Tt^z(x+\epsilon,x,\omega)^{-1}\,\Tt^z(x,T_{-\epsilon}\omega)\,\Tt^z(\epsilon,0,\omega)\;.
$$
Taking the limit $\epsilon\downarrow 0$ gives $\Tt^z(x,\omega)=\Tt^z(x,T_{0-}\omega)$ which implies
$M^z_\pm(T_{0-}\omega)=M^z_\pm(\omega)$. Similarly, the limit $\epsilon\downarrow 0$ of
$$
\Tt^z(x,\omega)\;=\;\Tt^z(x,x-\epsilon,\omega)\,
\Tt^z(x,T_{\epsilon}\omega)\,\Tt^z(0,-\epsilon,\omega)^{-1}\;,
$$
leads to
$$
\Tt^z(x,\omega)\;=\;e^{\Jj\Vv_x}\,\Tt^z(x,T_{0+}\omega)\,(e^{\Jj\Vv_0})^{-1}\;.
$$
As the jump at $x$ does not effect the square-integrability in \eqref{eq-L2sol}, this implies that
$$
(e^{\Jj\Vv_0})^{-1}\,
\left(\begin{array}{c} \one \\ \pm M_\pm^z(\omega) \end{array}\right)\,N\;=\;
\left(\begin{array}{c} \one \\ \pm M_\pm^z(T_{0+}\omega) \end{array}\right)
\;,
$$
for some invertible $L\times L$ matrix $N$. The upper entry implies that $N=(D_0^*-B_0^*(\pm M^z_\pm(\omega)))^{-1}$, the lower one
\begin{equation}
\label{eq-Moebdirect0}
\pm M_\pm^z(T_{0+}\omega)\;=\;(-C_0^*\pm A_0^*M^z_\pm(\omega))\,(D_0^*\mp B_0^* M^z_\pm(\omega))^{-1}\;.
\end{equation}
This is precisely the equation claimed in (iv) in the case $x=0$.
(v) follows from \eqref{eq-alphabeta} and the last 4 identities.
For (vi) we use $\Tt^z(x,\omega)=e^{\Jj\Vv_x}\Tt^z(x-,\omega)$ for $x>0$, giving
$$
\Phi^z_\pm(x,\omega)\;=\;e^{\Jj\Vv_x}\;\Phi^z_\pm(x-,\omega)\;=\;
e^{\Jj\Vv_x}\;\left(\begin{array}{c} \one \\ \pm M_\pm^z(T_{-(x-)}\omega) \end{array}\right)\,
\alpha^z_\pm(x-,\omega)\;,
$$
where (iii) was used in the second equality. The upper entry of this identity gives the first equality of (vi). The second one follows by replacing \eqref{eq-Moebdirect0} and using $A_xD_x^*-B_xC_x^*=\one$ and $A_xB_x^*=B_xA_x^*$. The following calculation gives (vii):
\begin{eqnarray*}
\partial_x\alpha^z_\pm(x,\omega)
& = &
(\one\;0)\;\partial_x\Tt^z(x,\omega)\;
\left(\begin{array}{c} \one \\ \pm M_\pm^z(\omega) \end{array}\right) \\
& = &
(0\;\one)\;\bigl(z-\Ww(T_{-x}\omega)\bigr)\;
\left(\begin{array}{c} \one \\ \pm M_\pm^z(T_{-x}\omega) \end{array}\right)
\;\alpha^z_\pm(x,\omega)\;.
\end{eqnarray*}
Finally,
\begin{eqnarray*}
\partial_x M^z_\pm(T_{-x}\omega)
\!\! & = & \!\!
(0\;\one)\;\partial_x\left[\Tt^z(x,\omega)\;
\left(\begin{array}{c} \one \\ \pm M_\pm^z(\omega) \end{array}\right)\,\alpha^z_\pm(x,\omega)^{-1}\right] \\
\!\! & = & \!\!
(\one\;0)\,\bigl(\Ww(T_{-x}\omega)-z\bigr)\,
\left(\begin{array}{c} \one \\ \pm M_\pm^z(T_{-x}\omega) \end{array}\right)
\mp M_\pm^z(T_{-x}\omega) \partial_x\alpha^z_\pm(x,\omega)\alpha^z_\pm(x,\omega)^{-1}
\,,
\end{eqnarray*}
so taking (vii) into account gives (viii).
\hfill $\Box$

\vspace{.2cm}

The fact that \eqref{eq-Moebdirect0} is a M\"obius transformation with a matrix out of SP$(2L,\CC)$ has a number of consequences which we regroup for later use.

\begin{coro}
\label{coro-moeb}
Let $x\in\Pp$ and set $M^z_\pm=M^z_\pm(T_{-x}\omega)$, $M^z_\pm(+)=M^z_\pm(T_{-x+}\omega)$ and $\Vv=\Vv_x$. Then

\vspace{.1cm}

\noindent {\rm (i)}
$\pm\, M_\pm^z(+)\,=\,(-C^*\pm A^*M^z_\pm)\,(D^*\mp B^* M^z_\pm)^{-1}$

\vspace{.1cm}

\noindent {\rm (ii)}
$ \pm\, M_\pm^z(+)\,=\,(\pm M^z_\pm B -D)^{-1}\,(C\mp M^z_\pm A)$

\vspace{.1cm}

\noindent {\rm (iii)}
$ M^z_+(+)+M^z_-(+)
\, = \, (D-M^z_+B)^{-1}\,(M^z_++M^z_-)\,(D^*+B^*M^z_-)^{-1}$\\
\noindent \phantom{{\rm (iii)} $ M^z_+(+)+M^z_-(+)$}
$\,=\,(D+M^z_-B)^{-1}\,(M^z_++M^z_-)\,(D^*-B^*M^z_+)^{-1} $

\vspace{.1cm}

\noindent {\rm (iv)}
$\pm M_\pm^z\;=\;(A\pm B M^z_\pm(+))\,(C\pm D M^z_\pm(+))^{-1}$

\vspace{.1cm}

\noindent {\rm (v)} $A\pm B M^z_\pm(+)=(D^*\mp B^*M^z_\pm)^{-1}$

\vspace{.1cm}

\noindent {\rm (vi)} $\Im m(M_\pm^z(+))\,=\,
(D\mp M^z_\pm B)^{-1}
\,\Im m(M^z_\pm)\,
\bigl((D\mp M^z_\pm B)^{-1}\bigr)^*
$

\vspace{.1cm}

\noindent {\rm (vii)} $\Im m(M_\pm^z(+))\,=\,
\bigl((D^*\mp B^*M^z_\pm)^{-1}\bigr)^*
\,\Im m(M^z_\pm)\,
(D^*\mp B^*M^z_\pm)^{-1}
$

\vspace{.1cm}

\noindent {\rm (viii)}
$\dot{M}^z_+(+)-\dot{M}^z_-(+)\,=\,
(M^z_+B-D)^{-1}\dot{M}^z_+(B^*M^z_+-D^*)^{-1}-
(M^z_-B+D)^{-1}\dot{M}^z_-(D^*+B^*M^z_-)^{-1}$

\end{coro}

\noindent {\bf Proof.} All this follows by short calculations using {\it e.g.} the Appendix of \cite{SB2} and the identities $AB^*=BA^*$, $CD^*=DC^*$ and $AD^*-BC^*=\one$.
\hfill $\Box$

\vspace{.2cm}

Now let us recall the definition of the Lyapunov exponents and state some of their properties.
Because $\Tt^z(x,\omega)$ is a cocycle by Lemma~\ref{lem-Ricatti},
Osceledec's theorem (see \cite{KS} for a concise statement) associates $2L$ Lyapunov exponents at $+\infty$ and $-\infty$ which will respectively be denoted by
$\gamma^z_{1}\geq\ldots\geq\gamma^z_{2L}$ and
$\hat{\gamma}^z_{1}\geq\ldots\geq\hat{\gamma}^z_{2L}$. 
Similarly, $\alpha_\pm^z(x,\omega)$ are other cocycles of $L\times L$
matrices, so again each has $L$ Lyapunov exponents at $+\infty$ and
$-\infty$ denoted by
$\gamma^{z,\pm}_{1}\geq\ldots\geq\gamma^{z,\pm}_{L}$ and
$\hat{\gamma}^{z,\pm}_{1}\geq\ldots\geq\hat{\gamma}^{z,\pm}_{L}$. Part
of the following proposition is copied from \cite{KS} (even though the
definition of $\gamma^{z,+}_{l}$ differs by a sign).

\begin{proposi}
\label{prop-Lyapprop} The various Lyapunov exponents satisfy:

\vspace{.1cm}

\noindent {\rm (i)} $\gamma^z_{l}=-\hat{\gamma}^z_{2L-l+1}$ for $l=1,\ldots,2L$

\vspace{.1cm}

\noindent {\rm (ii)} $\gamma^{z,\pm}_{l}=-\hat{\gamma}^{z,\pm}_{L-l+1}$ for $l=1,\ldots,L$

\vspace{.1cm}

\noindent {\rm (iii)} $\gamma^z_{l}=\gamma^{z,-}_l$ for $l=1,\ldots,L$ and $z\in\CC/\RR$

\vspace{.1cm}

\noindent {\rm (iv)} $\gamma^z_{l}=\gamma^{z,+}_{l-L}$ for $l=L+1,\ldots,2L$
and $z\in\CC/\RR$

\vspace{.1cm}

\noindent {\rm (v)} $\gamma^{\overline{z}}_{l}=-\gamma^{z}_{2L-l+1}$ for $l=1,\ldots,2L$

\vspace{.1cm}

\noindent {\rm (vi)} $\gamma^{z,+}_{l}=-\gamma^{\overline{z},-}_{L-l+1}$ for $l=1,\ldots,L$
and $z\in\CC/\RR$

\end{proposi}

\noindent {\bf Proof.} Items (i) and (ii) follow immediately from Lemma~5.2 of \cite{KS}. The other items can be proved as in Lemma~5.3 of \cite{KS} if one, moreover, uses the identity $\Tt^z(x,\omega)^{-1}=\Jj^* \Tt^{\overline{z}}(x,\omega)\Jj$ following from Wronskian identity \eqref{eq-Wronski} and invokes Corollary~\ref{coro-Mbounds} to show that $M^z_\pm(\omega)$ is uniformly bounded in $\omega$ for every fixed $z$.
\hfill $\Box$

\section{Kotani theory}
\label{sec-Kotani}

Kotani theory links the absolutely continuous spectrum of stochastic quasi-one-dimensional operators to the set of energies with vanishing Lyapunov exponents, by using analyticity arguments based on a few crucial identities. In all this section it is not needed that the stochastic Dirac operator has time-reversal symmetry or is of the particular random form given in \eqref{eq-randommodel}. Kotani theory for stochastic Dirac operators with bounded potentials was developed in \cite{Sun} by providing the relevant identities and then following closely the arguments of \cite{KS}. As already mentioned, the paper by Sun has some obvious errors which are corrected below. Moreover, we extend the theory in order to include singular potentials and potentials which may be complex-valued matrices. The singular potentials model a discrete version of Dirac operators (a satisfactory discrete analog acting on $\ell^2(\ZZ,\CC^{2L})$ does not exist).

\vspace{.2cm}

\begin{theo}
\label{theo-Kotanimain}
Let be given a stochastic family of Dirac operators with integrable and singular potentials. Then, for $k=1,\ldots,L$, the disjoint sets
$$
S_k\;=\;\left\{E\in\RR\,\left|\,\mbox{\rm exactly }2k\;\mbox{\rm Lyapunov exponents vanish at }E\;
\right.\right\}
$$
are an essential support of the absolutely continuous spectrum of multiplicity $2k$.
\end{theo}

Just as the crucial identities are different for discrete and continuous Schr\"odinger operators (compare \cite{KS}), there are some variations in the formulas in \cite{Sun} for stochastic Dirac operators with singular potentials as well.
We need to introduce further notations in order to state them.
Averaging over $\omega$ w.r.t. $\PP$ is denoted by $\EE$. Another average along the orbit of singular points is
$$
\EE_\Pp(f)\;=\;\EE\left( \lim_{x\to\infty}\frac{1}{x}\sum_{y\in\Pp\cap[0,x]}f(T_{-y}\omega)
\right)\;=\;\rho_\Pp\;\EE\left( \lim_{J\to\infty}\frac{1}{J}\sum_{j=1}^{J}f(T_{-x_j}\omega)
\right)\;,
$$
namely one first averages over the random sites of the singular potential. Note that $\EE_\Pp(1)=\rho_\Pp$ and that the average $\EE$ can be dropped $\PP$-almost surely. Furthermore, if $x_\Pp\in\Pp$ is the point in closest to the origin, then $T_{-x_\Pp}\omega$ has a singular point at the origin and $\EE_\Pp(f)=\int\PP(d\omega)\,f(T_{-x_\Pp}\omega)$. Hence $\EE_\Pp$ is closely linked to the Palm measure. Further the sum of the Lyapunov exponents is denoted by $\gamma^z=\sum_{l=1}^L\gamma^z_l$ and
we introduce two functions on $\CC/\RR$ by
$$
w^z_+\;=\;-\,\EE_\Pp\,\ln\left(\det(D-M^z_+B)\right)\,-\,\EE\,\Tr\bigl(R+M^z_+(Q-z)\bigr)\;,
$$
and
$$
w^z_-\;=\;\EE_\Pp\,\ln\left(\det(D^*+B^*M^z_-)\right)\,-\,\EE\,\Tr\bigl(-R^*+M^z_-(Q-z)\bigr)\;.
$$
By Corollary~\ref{coro-Mbounds} the imaginary part of $M^z_\pm$ is uniformly bounded away from $0$ so that $w^z_\pm$ are well-defined. The branch of the logarithm is chosen in a continuous way in $z$ (for each $\omega$ separately) so that Theorem~\ref{theo-WT} then shows that $w^z_\pm$ is analytic. The choice of the branch is of no importance below.
Finally for any smooth function $f$ on $\Omega$ we define $\partial f(\omega)=\partial_x f(T_{-x}\omega)|_{x=0}$ if $0\notin\Pp$.

\begin{theo}
\label{theo-Kotani}
Let $\Im m(z)\neq 0$.

\vspace{.1cm}

\noindent {\rm (i)} There is a constant $c\in\RR$ such that $w^z_+=w^z_-+\imath\, c$

\vspace{.1cm}

\noindent {\rm (ii)} $\gamma^z=-\,\Re e(w_\pm^z)$

\vspace{.1cm}

\noindent {\rm (iii)} $\partial_z w^z=\EE\,\Tr(\widehat{G}^z)$

\vspace{.1cm}

\noindent {\rm (iv)}  $2\,\gamma^z=\Im m(z)\,\EE\,\Tr\bigl((\one+|M^z_\pm|^2)\,(\Im m(M^z_\pm))^{-1}\bigr)$

\end{theo}

\vspace{.2cm}

Items (ii) and (iii) combined provide a Thouless formula for stochastic Dirac operators.
The proof is based on a series of algebraic identities which we check first.

\begin{lemma}
\label{lem-derivatives}
Let $\Im m(z)\neq 0$. Away from singular points, the following identities hold.

\vspace{.1cm}

\noindent {\rm (i)} $\partial\, \Tr\bigl(\ln(M^z_++M^z_-)\bigr)=\Tr\bigl(R^*+R+(Q-z)(M^z_+-M_-^z)\bigr)$

\vspace{.1cm}

\noindent {\rm (ii)} $\partial\,\Tr\bigl( (M^z_++M^z_-)^{-1}(\partial_zM^z_+-\partial_zM^z_-)\bigr) =
2\,\Tr\bigl(\widehat{G}^z_0\bigr)+\partial_z\Tr\bigl((Q-z)(M_+^z+M^z_-)\bigr)$

\vspace{.1cm}

\noindent {\rm (iii)} $\pm\,\partial \,\Tr\bigl(\ln(\Im m(M^z_\pm))\bigr) = 2\,\Re e \,\bigl(\Tr(W^z_\pm
)\bigr)-\Im m(z)\,
\Tr\bigl((\one+|M^z_\pm|^2)\,(\Im m(M^z_\pm))^{-1}\bigr)$

\vspace{.1cm}

where $W^z_+=R+(Q-z)M^z_+$ and $W^z_-=-R^*+(Q-z)M^z_-$

\vspace{.1cm}

\noindent {\rm (iv)}  $\partial_x\bigl[\alpha^z_\pm(x,\omega)^*
\Im m(M^z_\pm(T_{-x}\omega))
\alpha^z_\pm(x,\omega)\bigr]=\mp\,\Im m(z)\,\alpha^z_\pm(x,\omega)^*
\bigl(\one+|M^z_\pm(T_{-x}\omega)|^2\bigr)
\alpha^z_\pm(x,\omega)$

\end{lemma}

\noindent {\bf Proof.} In the formulas below all functions have the argument $T_{-x}\omega$, and one may then set $x=0$. Using Lemma~\ref{lem-Ricatti}(viii), a short calculation shows
\begin{equation}
\label{eq-deriid}
\partial\, \bigl(M^z_++M^z_-\bigr)
\; = \;
\bigl(M^z_++M^z_-\bigr)
\bigl(R^*-(Q-z)M^z_-\bigr)
\,+\,
\bigl(R+M^z_+(Q-z)\bigr)
\bigl(M^z_++M^z_-\bigr)
\;.
\end{equation}
Multiplying this by $(M^z_++M^z_-)^{-1}$ and then using the cyclicity of the trace shows the formula of (i).
For (ii), let us take the derivative $\partial_z$ of the Ricatti equation of Lemma~\ref{lem-Ricatti}(viii):
\begin{eqnarray*}
\partial\, \bigl(\dot{M}^z_+-\dot{M}^z_-\bigr)
& = &
-\,(2+({M}^z_+)^2+({M}^z_-)^2)\,+\,
\bigl(\dot{M}^z_+-\dot{M}^z_-\bigr)R^*\,+\,R\bigl(\dot{M}^z_+-\dot{M}^z_-\bigr)
\\
& & \;\,+\,
\dot{M}^z_+(Q-z){M}^z_+\,+\,
{M}^z_+(Q-z)\dot{M}^z_+\,+\,
\dot{M}^z_-(Q-z){M}^z_-\,+\,
{M}^z_-(Q-z)\dot{M}^z_-\;.
\end{eqnarray*}
Using this and \eqref{eq-deriid}, some algebra  directly leads to (ii) if one also uses the identity
$$
\Tr\bigl(\widehat{G}^z_0\bigr)
\;=\;
\Tr
( [(M^z_+)^{-1}+(M^z_-)^{-1}]^{-1}-(M^z_++M^z_-)^{-1})\;,
$$
following from Propositions~\ref{prop-Greenwithout} and \ref{prop-GreenHerg}.

\vspace{.1cm}

Next we turn to the proof of (iii). Let us set $M^z_\pm=X^z_\pm+\imath\,Y^z_\pm$ with $Y^z_\pm=\Im m (M^z_\pm)$. From $M^{\overline{z}}_\pm=(M^z_\pm)^*$ follows $X^z_\pm=X^{\overline{z}}_\pm=(X^z_\pm)^*$ and $Y^z_\pm=-Y^{\overline{z}}_\pm=(Y^z_\pm)^*$. Straightforward calculation then shows
$$
\partial\, Y^z_\pm
\,=\,
RY^z_\pm+Y^z_\pm R^*\pm X^z_\pm(Q-\Re e(z)) Y^z_\pm
\pm Y^z_\pm(Q-\Re e(z)) X^z_\pm
\mp \Im m(z)(\one+(X^z_\pm)^2-(Y^z_\pm)^2)
\,.
$$
Thus
\begin{eqnarray*}
\partial \,\Tr(\ln(Y^z_\pm))
& = &
\Tr\bigl(
R+R^*\pm 2\,X^z_\pm(Q-\Re e(z))\bigr)
\;\mp\; \Im m(z)\,\Tr\left(
(Y^z_\pm)^{-1}(\one+(X^z_\pm)^2-(Y^z_\pm)^2)\right)
\\
& = &
\pm\,2\,\Re e\;\bigl(\Tr(W^z_\pm)\bigr)\;\mp\;\Im m(z)\,
\Tr\left(
(Y^z_\pm)^{-1}(\one+(X^z_\pm)^2-(Y^z_\pm)^2)+2\,Y^z_\pm\right)
\\
& = &
\pm\,2\,\Re e\;\bigl(\Tr(W^z_\pm)\bigr)\;\mp\;\Im m(z)\,
\Tr\left(
(Y^z_\pm)^{-1}(\one+|X^z_\pm+\imath Y^z_\pm|^2)\right)\;,
\end{eqnarray*}
where in the last step we used $\Tr(Y^{-1}[X,Y])=0$. Finally let us consider (iv). When calculating the derivative on the l.h.s. the product rule leads to three terms. The term containing $\partial\, Y^z_\pm$ is given by the above formula, those involving derivatives of $\alpha^z_\pm(x,\omega)$ by Lemma~\ref{lem-Ricatti}(vii). Hence it is sufficient to check
$$
\bigl(-R^*\mp(Q-z)M^z_\pm\bigr)^*Y^z_\pm\;+\;\partial\, Y^z_\pm\;+\;
Y^z_\pm\bigl(-R^*\mp(Q-z)M^z_\pm\bigr)
\;= \;\mp\,\Im m(z)\;\bigl(\one+|M^z_\pm|^2\bigr)\;.
$$
Again this follows from some algebra.
\hfill $\Box$

\vspace{.2cm}

\noindent {\bf Proof} of Theorem~\ref{theo-Kotani}. (i) Set $I^z=\EE\;\Tr\bigl(R^*+R+(Q-z)(M^z_+-M^z_-)\bigr)$. By the ergodic theorem and Lemma~\ref{lem-derivatives}(i), $\PP$-almost surely
\begin{eqnarray*}
I^z
& = &
\lim_{y\to\infty}\,\frac{1}{y}\,\int^y_0 dx\;
\Tr\bigl(R^*(T_{x}\omega)+R(T_{x}\omega)+(Q(T_{x}\omega)-z)(M^z_+(T_{x}\omega)-M^z_-(T_{x}\omega))\bigr)
\\
& = &
\lim_{y\to\infty}\,\frac{1}{y}\,\int^0_{-y} dx\;
\partial_x\,\Tr\bigl(\ln (M^z_+(T_{-x}\omega)+M^z_-(T_{-x}\omega))\bigr)
\\
& = &
\lim_{y\to\infty}\,\frac{1}{y}\,\sum_{-y\leq x_j\leq 0}
\left[\,\ln\bigl(\det (M^z_+(T_{-x}\omega)+M^z_-(T_{-x}\omega))\bigr)
\bigr|^{x_j-}_{x_{j-1}+}\;+\;2\pi\imath\, n_j\,\right]
\;,
\end{eqnarray*}
where $\Pp=(x_j)_{j\in\ZZ}$ with $x_{j-1}\leq x_j$ and $n_j\in\ZZ$ denotes the number of branches of the logarithm needed in the integral from $x_{j-1}$ to $x_j$ minus $1$. Now by Lemma~\ref{lem-Ricatti}(iv),
$M^z_\pm(T_{-(x_{j-1}+)}\omega)=M^z_\pm(T_{-x_{j-1}}\omega)$. On the other hand, we calculate
$M^z_+(T_{-x_{j}+}\omega)+M^z_-(T_{-x_{j}+}\omega)$ by Corollary~\ref{coro-moeb}(iii). Thus regrouping the terms shows that
$$
I^z\,=\,
\lim_{y\to\infty}\,\frac{1}{y}\,\sum_{-y\leq x_j\leq 0}
\left[-\ln\bigl(\det (D_{j}-M^z_+(T_{-x_j}\omega)B_{j}\bigr)
-\ln\bigl(\det (D^*_{j}+B^*_{j}M^z_-(T_{-x_j}\omega))\bigr)
\,+\,2\pi\imath\, n_j\,\right]\,.
$$
Hence if $c$ is the average of $2\pi n_j$ over $\Pp$, we have shown
$$
I^z\,=\,
-\,\EE_\Pp\,\ln\bigl(\det (D-M^z_+B)\bigr)
\,-\,\EE_\Pp\,\ln\bigl(\det (D^*+B^*M^z_-)\bigr)
\,+\,\imath\, c\;,
$$
and thus (i). For (ii) let us start from a formula for $\gamma^z$ which follows from the identities 
stated in Proposition~\ref{prop-Lyapprop}:
$$
\gamma^z\;=\;
\lim_{y\to\infty}\,\frac{1}{y}\,\ln
\bigl(|\det(\alpha^{{z}}_-(y,\omega))|\bigr)
\;,
$$
where the convergence holds $\PP$-almost surely. Telescoping and regrouping gives
\begin{eqnarray*}
\gamma^z & = &
\lim_{y\to\infty}\,\frac{1}{y}\sum_{0<x_j<y}\left[
\;\ln
\bigl(|\det(\alpha^{{z}}_-(x_j+,\omega))|\bigr)
-
\ln
\bigl(|\det(\alpha^{{z}}_-(x_{j-1}+,\omega))|\bigr)\;\right]
\\
& = &
\lim_{y\to\infty}\,\frac{1}{y}\sum_{0<x_j<y}\left[
\ln
\bigl(|\det(\alpha^{{z}}_-(x_j+,\omega)\alpha^{{z}}_-(x_j-,\omega)^{-1})|\bigr)
+\int^{x_j}_{x_{j-1}} \!dx\,
\partial_x\ln
\bigl(|\det(\alpha^{{z}}_-(x,\omega))|\bigr)
\right]
.
\end{eqnarray*}
The first contribution can be evaluated with Lemma~\ref{lem-Ricatti}(vi) and the definition of $\EE_\Pp$, the second contribution be summed up and the integrand evaluated:
$$
\gamma^z\;=\;-\,
\EE_\Pp\,\ln
\bigl(|\det(D^*+B^*M^z_-)|\bigr)
\;+\;\Re e\,\lim_{y\to\infty}\,\frac{1}{y}
\int^{y}_{0} \!dx\,
\Tr\bigl(\alpha^{{z}}_-(x,\omega)^{-1}\partial_x\alpha^{{z}}_-(x,\omega)\bigr)
\,.
$$
Finally the last expression can be calculated using Lemma~\ref{lem-Ricatti}(vii) and then the ergodic theorem completes the proof of (ii). Let us point out that one could have started from
$$
\gamma^z\;=\;-\,\lim_{y\to\infty}\,\frac{1}{y}\,\ln \bigl(|\det(\alpha^{\overline{z}}_+(y,\omega))|\bigr)
\;.
$$
Then a similar calculation leads to $\gamma^z=-\,\Re e(w^z_+)$. Because $w^z_\pm$ are analytic, this also provides an alternative proof of (i).

\vspace{.1cm}

(iii) Let us set $J^z=2\,\EE\,\Tr\bigl(\widehat{G}^z\bigr)+\partial_z\EE\,\Tr\bigl(R-R^*+(Q-z)(M_+^z+M^z_-)\bigr)$.
Because the probability of having a singular potential at $0$ vanishes,
$\EE\,\Tr\bigl(\widehat{G}^z\bigr)$ can be replaced by $\EE\,\Tr\bigl(\widehat{G}^z_0\bigr)$. Furthermore
the term $R-R^*$ drops out due to the derivative $\partial_z$. Hence Lemma~\ref{lem-derivatives}(ii), the ergodic theorem and reordering of the terms imply as above that $\PP$-almost surely
$$
J^z\;=\;
\lim_{y\to\infty}\,\frac{1}{y}\,\sum_{-y\leq x_j\leq 0}
\Tr\left((M^z_+(T_{-x}\omega)+M^z_-(T_{-x}\omega))^{-1}
(\dot{M}^z_+(T_{-x}\omega)-\dot{M}^z_-(T_{-x}\omega))
\right)
\bigr|^{x_j-}_{x_{j}}
\;,
$$
where we also used the left-continuity of $x\in\RR\mapsto M^z_\pm(T_x\omega)$. The terms with $x_j-$ now have to be evaluated using Lemma~\ref{lem-Ricatti}(iv) or its equivalent formulations. The factor $(M^z_+(T_{-x_j+}\omega)+M^z_-(T_{-x_j+}\omega))^{-1}$ is given by
the inverse of Corollary~\ref{coro-moeb}(iii).
Corollary~\ref{coro-moeb}(viii) moreover allows to calculate
$\dot{M}^z_+(T_{-x_j+}\omega)-\dot{M}^z_-(T_{-x_j+}\omega)$ .
Replacing both identities then shows
%
%
\begin{eqnarray*}
J^z &  =  & \EE_\Pp\;\Tr\left(
(M^z_++M^z_-)^{-1}\left[\,(D+M^z_- B)(D-M^z_+ B)^{-1}\;\dot{M}^z_+
\right.\right.
\\
& & \;\;\;\;\;\;\;\;\;\;\;\;\;\;\;\;\;\;\;\;\;\;\;\;\;\;\;\;\;\;\;\;\;\;\;\;\left.\left.
+
\;\dot{M}^z_-\;(D^*+B^* M^z_-)^{-1}(B^* M^z_+-D^*)
\;-\;\dot{M}^z_+\;+\;\dot{M}^z_-\right]
\right)
\\
& = & \partial_z\;\EE_\Pp\,
\left[\,\Tr\bigl(\ln(D^*+B^*M^z_-)\bigr)\,-\,\Tr\bigl(\ln(D-M^z_+ B)\bigr)\right]
\;.
\end{eqnarray*}
Due to the definitions of $J^z$ and $w^z_\pm$ this concludes the proof of (iii).

(iv) We set $K^z_\pm=\EE\bigl(2\,\Re e\bigl(\Tr(W^z_\pm)\bigr)-\Im m(z)\,\Tr ((\one+|M^z_\pm|^2)\Im m(M^z_\pm)^{-1})\bigr)$. By Lemma~\ref{lem-derivatives}(iii) and the ergodic theorem one has $\PP$-almost surely
$$
\pm\,K^z_\pm\;=\;
\lim_{y\to\infty}\,\frac{1}{y}\,\sum_{-y\leq x_j\leq 0}
\ln\left(\det(\Im m(M^z_\pm(T_{-x}\omega)))
\right)
\bigr|^{x_j-}_{x_{j}}
\;.
$$
Now evaluate $\Im m(M^z_\pm(T_{-x_j+}\omega))$ by Corollary~\ref{coro-moeb}(vi). This implies
$$
K^z_\pm\;=\;\mp\,\EE_\Pp\,\ln\bigl(\det(|D\mp M^z_\pm B|^2)\bigr)\;=\; \mp\,2\,\Re e\,\EE_\Pp\,\ln\bigl(\det(D\mp M^z_\pm B)\bigr)\;.
$$
Similarly, using Corollary~\ref{coro-moeb}(vii), $K^z_\pm=\mp\,2 \,\Re e\,\EE_\Pp\,\ln\bigl(\det(D^*\mp B^*M^z_\pm)\bigr)$. From these identities one readily completes the proof.
\hfill $\Box$

\vspace{.2cm}

The second part of the following theorem establishes Theorem 6.6 of \cite{KS} also for complex valued potentials. 

\begin{theo}
\label{theo-partialsum}
Consider the positive operator
$U^z_\pm=(\Im m(M^z_\pm))^{\frac{1}{2}}(\one+|M^z_\pm|^2)^{-1}(\Im m(M^z_\pm))^{\frac{1}{2}}$ and denote its eigenvalues by
$u^z_{1,\pm}\geq\ldots\geq u^z_{L,\pm}\geq 0$. Further let $E\in\RR$, $\epsilon>0$ and $k=1,\ldots,L$. Then
\begin{equation}
\label{eq-ineg}
\EE\;\sum_{l=1}^k\,\frac{1}{u^{E+\imath \epsilon}_{l,\pm}}\;\leq\;\frac{2}{\epsilon}\;
\sum_{l=1}^k\,\gamma^{E\mp\imath\epsilon}_{L+1-l}\;.
\end{equation}
If furthermore $E$ is such that $\gamma^E_l=\lim_{\epsilon\to 0}\gamma^{E+\imath\epsilon}_l$ exists for $l=1,\ldots,L$, then
\begin{equation}
\label{eq-ineg2}
\EE\;\sum_{l=1}^k\,\frac{1}{u^{E+\imath \epsilon}_{l,\pm}}\;\leq\;\frac{2}{\epsilon}\;
\left[\sum_{l=1}^k\,\gamma^{E+\imath\epsilon}_{L+1-l}\;+\;
\sum_{l=k+1}^L\,(\gamma^{E+\imath\epsilon}_{L+1-,}-
\gamma^{E}_{L+1-l})
\right]\;.
\end{equation}
\end{theo}

\noindent {\bf Proof.} This is an adaption and slight generalization of the proof of Theorems 6.5 and 6.6 of \cite{KS} (the reasoning in \cite{Sun} is erroneous at several points). For any $L\times L$  matrix $F$ let $\Lambda^k F$ and $d\Lambda^k F$ the second quantizations on the fermionic tensor product $\Lambda^k\CC^L$, such that $e^{d\Lambda^k F}=\Lambda^k e^{F}$. Let $z=E+\imath\epsilon$ and $Y^z_\pm=\Im m(M^z_\pm)$. Define $F^z_\pm(x,\omega)=Y^z_\pm(T_{-x}\omega)^{\frac{1}{2}}\alpha^z_\pm(x,\omega)$. Then
$$
\partial_x\,\Lambda^{k}|F^z_\pm(x)|^2 \; = \;
\Lambda^{k}F^z_\pm(x)^*
\left( d\Lambda^{k}(F^z_\pm(x)^{-1})^*\;\partial_x |F^z_\pm(x)|^2\; F^z_\pm(x)^{-1}\right)
\Lambda^{k}F^z_\pm(x)\;.
$$
Thus by Lemma~\ref{lem-derivatives}(iv)
$$
\partial_x\,\Lambda^{k}|F^z_\pm(x,\omega)|^2 \;= \;
\mp\;\Im m(z)\;\Lambda^{k}F^z_\pm(x,\omega)^*
\left( d\Lambda^{k} U^z_\pm(T_{-x}\omega)^{-1}\right)
\Lambda^{k}F^z_\pm(x,\omega)\;,
$$
so that for $\Im m(z)>0$
\begin{eqnarray*}
\partial_x\,\Lambda^{k}|F^z_+(x,\omega)|^2  & \geq &
-\;\Im m(z)\;\|d\Lambda^{k}U^z_+(T_{-x}\omega)^{-1}\|\;
\Lambda^{k}|F^z_+(x,\omega)|^2\;,
\\
\partial_x\,\Lambda^{k}|F^z_-(x,\omega)|^2 & \leq &
\Im m(z)\;\|d\Lambda^{k}U^z_-(T_{-x}\omega)^{-1}\|\;
\Lambda^{k}|F^z_-(x,\omega)|^2\;.
\end{eqnarray*}
Integrating hence gives
\begin{eqnarray*}
\Lambda^{k}|F^z_+(x,\omega)|^2  & \geq &
\exp\left(-\;\Im m(z)\;\int^x_0dy\,\|d\Lambda^{k}U^z_+(T_{-y}\omega)^{-1}\|\right)
\Lambda^{k}|F^z_+(0,\omega)|^2\;,
\\
\Lambda^{k}|F^z_-(x,\omega)|^2 & \leq &
\exp\left(\Im m(z)\;\int^x_0dy\,\|d\Lambda^{k}U^z_-(T_{-y}\omega)^{-1}\|\right)
\Lambda^{k}|F^z_-(0,\omega)|^2\;.
\end{eqnarray*}
Note that by Lemma~\ref{lem-Ricatti}(vi) and Corollary~\ref{coro-moeb}(vii) the functions
$|F^z_{\pm}(x,\omega)|$ are actually smooth also for $x\in\Pp$.
We combine this with the inequalities
$$
\|\Lambda^{k}\alpha^z_\pm(x,\omega)\|^2\,\|\Lambda^{k}Y^z_\pm(T_{-x}\omega)^{-1}\|^{-1}\;\leq\;
\|\Lambda^{k}|F^z_\pm(x,\omega)|^2\|\;
\leq\;
\|\Lambda^{k}\alpha^z_\pm(x,\omega)\|^2\,\|\Lambda^{k}Y^z_\pm(T_{-x}\omega)\|\;.
$$
Taking logarithms thus shows
$$
\ln\left(\frac{\|\Lambda^{k}\alpha^z_-(x,\omega)\|^2}{\|\Lambda^{k}Y^z_-(T_{-x}\omega)^{-1}\|} \right)
\;\leq\;\Im m(z)\int^x_0dy\,\|d\Lambda^{k}U^z_+(T_{-y}\omega)^{-1}\|\,+\,
\ln\left(\|\Lambda^{k}|F^z_+(0,\omega)|^2 \|\right)\;.
$$
Now by Corollary~\ref{coro-Mbounds}, $Y^z_-(\omega)^{-1}$ is uniformly bounded in $\omega$. Thus dividing by $x$ and then taking the limit $x\to\infty$ shows by Proposition~\ref{prop-Lyapprop}(iii) and the ergodic theorem
\begin{equation}
\label{eq-bound-}
2\,\sum_{l=1}^k\gamma^z_l\;\leq\;\Im m(z)\;\EE\,\|
d\Lambda^{k}(U^z_-)^{-1}\|\;=\;\Im m(z)\;\EE\,\sum_{l=L-k+1}^L\,\frac{1}{u_{-,l}^z}\;.
\end{equation}
Combining this with $k$ replaced by $L-k$ together with Theorem~\ref{theo-Kotani}(iv) stating
$$
2\,\sum_{l=1}^L\gamma^z_l\;=\;\Im m(z)\;\EE\,\sum_{l=1}^L\,\frac{1}{u_{\pm,l}^z}\;,
$$
proves inequality \eqref{eq-ineg} for the sign $-$. Similarly one has
$$
\ln\left(\|\Lambda^{k}\alpha^z_+(x,\omega)\|^2 \right)
\;\geq\;-\,\Im m(z)\int^x_0dy\,\|d\Lambda^{k}U^z_+(T_{-y}\omega)^{-1}\|\,+\,
\ln\left(\frac{\|\Lambda^{k}|F^z_+(0,\omega)|^2\|}{\|\Lambda^{k}Y^z_+(T_{-x}\omega)\|}\right)
\;.
$$
As the last term is bounded along the orbit, Proposition~\ref{prop-Lyapprop}(iii) now implies
\begin{equation}
\label{eq-bound+}
2\,\sum_{l=1}^k\gamma^{\overline{z}}_l\;\leq\;
\Im m(z)\;\EE\,\sum_{l=L-k+1}^L\,\frac{1}{u_{+,l}^z}\;,
\end{equation}
which again combined with Theorem~\ref{theo-Kotani}(iv) proves \eqref{eq-ineg} for the sign $+$.

\vspace{.1cm}

For the proof of \eqref{eq-ineg2} we need the following general fact. If $T,S>0$ are two positive matrices, then the positive operators $T^{\frac{1}{2}}ST^{\frac{1}{2}}$ and $S^{\frac{1}{2}}TS^{\frac{1}{2}}$ have the same spectrum (this follows from $\Tr\bigl((T^{\frac{1}{2}}ST^{\frac{1}{2}})^n\bigr)=
\Tr\bigl((S^{\frac{1}{2}}TS^{\frac{1}{2}})^n\bigr)$ for all $n\in\NN$). Hence $u^z_{\pm,k}$ are also the eigenvalues of the imaginary part of the Herglotz function
$(\one+|M^z_\pm|^2)^{-\frac{1}{2}}M^z_\pm(\one+|M^z_\pm|^2)^{-\frac{1}{2}}$ and by the Herglotz representation theorem it follows as in \cite{KS} that
$$
\frac{\epsilon}{u^{E+\imath\epsilon}_{\pm,k}}\;\geq\;
\frac{\delta}{u^{E+\imath\delta}_{\pm,k}}
\qquad \mbox{ for }\epsilon\geq\delta>0\;.
$$
Combining this fact with Theorem~\ref{theo-Kotani}(iv) and the bounds \eqref{eq-bound-} and \eqref{eq-bound+} gives
$$
\EE\,\sum_{l=1}^k\,\frac{\epsilon}{u_{\pm,l}^{E+\imath\epsilon}}
\; \leq \;
\EE\,\sum_{l=1}^L\,\frac{\epsilon}{u_{\pm,l}^{E+\imath\epsilon}}
\;-\;\EE\,\sum_{l=k+1}^L\,\frac{\delta}{u_{\pm,l}^{E+\imath\delta}}
\;\leq \;
2\,\sum_{l=1}^{L}\,\gamma^{E+\imath\epsilon}_{l}\;-\;
2\,\sum_{l=1}^{L-k}\,\gamma^{E\mp\imath\delta}_{l}
\;.
$$
Now taking the limit $\delta\to 0$ leads to \eqref{eq-ineg2}.
\hfill $\Box$

\vspace{.2cm}

From this point on the proof of Theorem~\ref{theo-Kotanimain} is line by line the same as in \cite{KS}.

\section{Time reversal symmetry and Coupling Hypothesis}
\label{sec-TRI}

None of the results of Sections~\ref{sec-WT} to \ref{sec-Kotani} used
the time reversal invariance \eqref{eq-TRS}. In this section, we first
implement this symmetry and then describe the model
of Theorem~\ref{theo-main} in more detail and state the Coupling
Hypothesis. The proof of the following result is immediate.

\begin{proposi}
\label{prop-TRS} Suppose that $H$ is time-reversal invariant, namely satisfies {\rm \eqref{eq-TRS}}.
Then
$$
\Jj^*\overline{\Tt^z(x)}\Jj\;=\;\Tt^{\overline{z}}(x)\;,\qquad
\overline{M^z_\pm}\;=\;-(M^{\overline{z}}_\pm)^{-1}\;.
$$
\end{proposi}

\begin{proposi}
\label{prop-GreenSymmetry}
If $H_\Vv$ has time-reversal symmetry, the averaged Green matrix satisfies
\begin{equation}
\label{eq-GreenSymmetry}
\Jj^*\widehat{G}_\Vv^z\Jj
\; =\;
(\widehat{G}_\Vv^z)^t\;,\qquad
\Jj^*\Im m(\widehat{G}_\Vv^{\overline{z}})\Jj
\; =\;
\overline{\Im m(\widehat{G}_\Vv^z)}\;.
\end{equation}
If furthermore $\phi=(v,\Jj \overline{v})$ for some $v\in\CC^{2L}$
satisfying $v^*\Jj \overline{v}=0$, then the $2\times 2$ matrix
$\phi^*\widehat{G}_\Vv^z \phi$ is a constant multiple of the
identity. 
\end{proposi}

\noindent {\bf Proof.} The Hamiltonian satisfies
$\Jj^*H_\Vv\Jj=\overline{H_\Vv}$ so that 
$\Jj^*(H_\Vv-z)^{-1}\Jj=\overline{(H_\Vv-\overline{z})^{-1}}$. This
implies that for any vectors $v,w\in\CC^{2L}$,
$v^*\Jj^*\widehat{G}_\Vv^z\Jj
w=\overline{w}^*\widehat{G}_\Vv^z\overline{v}= 
v^*(\widehat{G}_\Vv^z)^tw$ which implies the first identity in
\eqref{eq-GreenSymmetry}, from which the second one can be directly
deduced. As to the last point, for any vector $w$ one has
$w^*\widehat{G}_\Vv^z
w=w^t(\widehat{G}_\Vv^z)^t\overline{w}=w^t\Jj^*\widehat{G}_\Vv^z\Jj\overline{w}= 
(\Jj\overline{w})^*\widehat{G}_\Vv^z\Jj\overline{w}$. Moreover, for
any $w=\lambda v+\lambda'\Jj\overline{v}\in 
\;$Ran$(\phi)$, one checks the orthogonality $w^*\Jj
\overline{w}=0$. These facts imply $w^*\widehat{G}_\Vv^z
w=\frac{1}{2}\,\Tr(\phi^*\widehat{G}_\Vv^z \phi)\,\|w\|^2$. 
\hfill $\Box$

\vspace{.2cm}

The last statement of Proposition~\ref{prop-GreenSymmetry} reflects
Kramers' degeneracy stating that the spectrum of a time reversal
invariant Hamiltonian with odd spin has even multiplicity. In
particular, for eigenstates $H\psi=E\psi$ gives $H\Jj\overline{\psi} =
E\Jj\overline{\psi}$. For the same reason, the singular values of the
transfer matrices are degenerate (see Lemma~\ref{lemma-decomp}(ii))
which implies the degeneracy of the Lyapunov spectrum.

\vspace{.2cm}

Next let us come to the construction of the stochastic Dirac operators of Theorem~\ref{theo-main} and of the associated dynamical system. Let $s\in [0,1)=\RR/\ZZ$. Each operator $H(\omega)$ is of the form \eqref{eq-Dirac} with singular potentials at $\Pp=\ZZ+s$, hence $x_j=j+s$. The $\Vv_j$ are drawn independently and identically out of $\Jj\,$so$^*(2L)$ with some probability law $\pp_\Vv$ with compact support. Furthermore the  potential $\Ww\in L^1_\loc(\RR,\Jj\mbox{so}^*(2L))$ is of the form
\begin{equation}
\label{eq-randommodel}
\Ww(x)\;=\;\sum_{j\in\ZZ}\sum_{k=1}^K \lambda_{j,k}\,\Ww_k(x+s-j+1)\;,
\end{equation}
where $K\in\NN$, each $\Ww_k\in L^1_\loc(\RR,\Jj\mbox{so}^*(2L))$ has
support $[0,1]$ and the vectors
$(\lambda_{j,k})_{k=1,\ldots,K}\in\RR^K$ are also drawn independently
and identically according to a probability distribution $\pp_\Ww$ with
compact support. Then $\Omega$ is a compact subset of 
$(\Jj\,$so$^*(2L)\times\RR^K)^{\times\ZZ}\times \RR/\ZZ$ and
$\PP=(\pp_\Ww\times \pp_\Vv)^{\times\ZZ}\times ds$. The $\RR$-action
$T$ is the natural right shift on $\Omega$ and $\PP$ is indeed ergodic
and even mixing w.r.t. $T$. 
In order to state the main hypothesis on the randomness, it is
convenient to introduce the transfer matrix $\Tt^z(\Ww,\Vv)$ as the
solution $\Tt^z(1,0)$ of \eqref{eq-transfer} with potential $\Ww$ and
jump $e^{\Jj \Vv}$ at $1$. Setting
$\lambda_j=(\lambda_{j,k})_{k=1,\ldots,K}$ (which determines the
potential bump $\Ww_j=\sum_{k=1}^K \lambda_{j,k}\,\Ww_k$ between $j-1$
and $j$), this notation implies 
$\Tt^z(\lambda_j,\Vv_j)=\Tt^z(j+s,j+s-1,\omega)$
where the transfer matrix on the r.h.s. is defined by
\eqref{eq-transfer} with the Hamiltonian $H(\omega)$. 

\vspace{.2cm}

\noindent {\bf Coupling Hypothesis:} {\it The semi-group generated by $\{\Tt^E(\lambda,\Vv)\,|\,(\lambda,\Jj \Vv)\in\mbox{\rm supp}(\pp_\Ww\times \pp_\Vv)\,\}$ is Zariski dense in} SO$^*(2L)$ {\it for all $E\in\RR$. }

\vspace{.2cm}

Let us stress that this hypothesis can be verified if $\pp_\Ww\times
\pp_\Vv$ is supported on a finite set of points, and also if either
$\pp_\Ww$ or $\pp_\Vv$ is concentrated on a single point, notably the
disorder is given only by a random potential $\Ww$ or the random Dirac
peaks $\Vv_j\delta_j$. 
Furthermore this hypothesis is satisfied whenever the set of
$\Tt^E(\lambda,\Vv)$ contains an 
open set (this property does not depend on $E$). This is {\it e.g.}
the case if $\pp_\Vv$ contains an absolutely continuous part w.r.t. to
the Haar measure.

\section{The Lyapunov spectrum}
\label{sec-Lyap}

This section proves a criterion for the distinctness (apart from Kramers' degeneracy) of the Lyapunov exponents for random products of matrices in SO$^*(2L)$. It can be immediately applied to the transfer matrices if the Coupling Hypothesis holds. On the other hand, we believe it to be of somewhat independent interest and thus took care to make it readable without reference to the rest of the paper.
Instead of the group $\SO^*(2L)$ as defined in the introduction it will be more convenient to work with an isomorphic group $\GM$ for which the polar decomposition takes a more simple form. Thus we  define in case of even $L=2d$ and odd $L=2d+1$ respectively
$$
A
\;=\;
\frac{1}{\sqrt{2}} \left(\begin{matrix}
\one_d & \one_d \\ \imath \one_d & -\imath \one_d
\end{matrix}\right)
\;,\qquad
A
\;=\;
\frac{1}{\sqrt{2}} \left(\begin{matrix}
\one_d & 0 & \one_d \\
0 & \sqrt{2} & 0 \\
\imath \one_d & 0& -\imath \one_d
\end{matrix}\right)
\;,
$$
where $d\times d$ square matrices carry the index $d$. Then introduce
$\Aa=\mbox{diag}(A,A)$
which satisfies $\Aa^* = \Aa^{-1}$ and set $\GM=\Aa^* \SO^*(2L) \Aa$. 
This group consists
of all $2L\times 2L$ matrices $\Mm$ satisfying
\begin{equation}\label{eq-defG}
\Mm^*\Jj \Mm\;=\;\Jj\;,\qquad
\Mm^t \Ss  \Mm\;=\; \Ss\;,
\end{equation}
where $\Ss=\mbox{diag}(A^tA,A^tA)$.
Note that the matrices $\Jj$ and $\Ss$ commute, $\Jj^*=-\Jj =\Jj^{-1}$ and $\Ss^*=\Ss=\Ss^{-1}$.

\begin{lemma}
\label{lemma-decomp}
Let $\Mm\in\GM$ and $v \in \CC^{2L}$.

\vspace{.1cm}

\noindent {\rm (i)} $ \Mm^* \in \GM$

\vspace{.1cm}

\noindent {\rm (ii)} If $ \Mm v = \lambda v$, then
$ \Mm^*\Jj v = \lambda^{-1} \Jj v$, $ \Mm \Jj \Ss \overline{v} = \overline{\lambda} \Jj \Ss \overline{v}$ and
$ \Mm^* \Ss \overline{v} = \overline{\lambda}^{-1} \Ss \overline{v}$.

\vspace{.1cm}

\noindent {\rm (iii)} The vectors $v$ and $\Jj \Ss \overline{v}$ are linearly independent for $v\neq 0$.

\vspace{.1cm}

\noindent {\rm (iv)}
For $\Mm>0, \Mm\in \GM$, there exists $\Uu \in \GM \cap {\rm SU}(2L)$ such that
$\Uu\Mm \Uu^*=\Dd$, where

$\Dd=\diag(a_1,\ldots, a_d, 1, a_1^{-1},\ldots, a_d^{-1}, a_1^{-1},\ldots, a_d^{-1}, 1, a_1,\ldots, a_d)$
if $L=2d+1$ and

$\Dd=\diag(a_1,\ldots, a_d, a_1^{-1},\ldots, a_d^{-1}, a_1^{-1},\ldots, a_d^{-1}, a_1,\ldots, a_d)$
in case $L=2d$, with real constants

$a_1 \geq a_2 \geq \ldots a_d \geq 1$. Note that $\Dd\in \GM$.

\vspace{.1cm}

\noindent {\rm (v)} There are unitary matrices $\Kk,\Uu \in  \GM \cap {\rm SU}(2L)$ and a diagonal matrix $\Dd$ as in {\rm (iv)}
such that

$\Mm=\Kk\Dd\Uu$.

\vspace{.1cm}

\noindent {\rm (vi)} One has $\det ( \Mm) = 1$ and the group $\GM$ is connected.

\end{lemma}

\noindent{\bf Proof.} (i) follows by inverting the relations in \eqref{eq-defG}.
For (ii) note that $ \Mm^* \Jj  \Mm = \Jj$ implies $\Jj^*  \Mm^* \Jj =  \Mm^{-1}$. Hence
$\Jj^*  \Mm^* \Jj v = \lambda^{-1} v$ implies $ \Mm^* \Jj v = \lambda^{-1} v$.
\noindent From $ \Mm^t \Ss  \Mm = \Ss$ it follows that $\Ss  \Mm^t \Ss =  \Mm^{-1} = \Jj^*  \Mm^* \Jj$.
Taking the transpose one obtains $\Ss  \Mm \Ss = \Jj^* \overline{ \Mm} \Jj$ and hence
$\Ss  \Mm \Ss \Jj \overline{v} = - \overline{\lambda} \Jj^* \overline{v}$ and therefore $ \Mm \Ss \Jj \overline{v} = \overline{\lambda} \Ss \Jj\overline{v}$.
Now using the same calculation as above yields the last equation.

\vspace{.1cm}

\noindent (iii) Writing $v=\binom{a}{b}$ and $\Jj\Ss \overline{v} 
= \lambda v$ gives
$\lambda a = A^tA \overline{b}$ and $\lambda b = - A^t A \overline{a}$.
As $A^t A$ is real and $(A^t A)^2=\one$,
this implies $|\lambda|^2 a = A^t A \overline{\lambda} \overline{b} = -a$ and therefore
$(1+|\lambda|^2) a = 0$ implying $a=0$ and $b=0$ and hence $v=0$.
Therefore these vectors are linearly dependent if and only if $v=0$.\\
(iv)  First we need some basic facts. We say that a subspace $\VM$ of $\CC^{2L}$ is $\GM$-like
if for any vector $v \in \VM $ one has $\Jj v, \Ss \overline{v}, \Jj \Ss \overline{v} \in \VM$.
The space spanned by $v,\Jj v, \Ss \overline{v}$ and $\Jj \Ss \overline{v}$ is $\GM$-like.
The intersection of two $\GM$-like subspaces is $\GM$-like.
Furthermore, if $\VM$ is $\GM$-like, then also the orthogonal complement $\VM^\perp$ is $\GM$-like.
To see this, take $v \in \VM, w \in \VM^\perp$ then
$\langle \Ss \overline{w},v \rangle = \langle \overline{w}, \Ss v \rangle = \overline{\langle w, \Ss\overline{v} \rangle}=0$,
and $\langle \Jj w, v \rangle = - \langle w, \Jj v \rangle = 0$. Therefore
$\Ss \overline{w}, \Jj w \in \VM^\perp$ and hence also $\Jj\Ss \overline{w} \in \VM^\perp$.\\
For $\Mm>0$ the eigenspaces are orthogonal.
Let $\VM_1$ be the eigenspace for the value $1$
(possibly only the zero vector) and $\VM_0$ be the orthogonal complement.
By (ii) and the consideration above, these spaces are $\GM$-like and they are invariant under
$ \Mm^* \Mm$. By (ii) and (iii) the dimension of $\VM_0$ is divisible by 4, say ${\rm dim}\VM_0 = 4r$.

\vspace{.1cm}

\noindent {\bf First claim:}
$\VM_0$ has an orthonormal basis of eigenvectors of $ \Mm^*  \Mm$ of the form \\
$v_1, v_2,\ldots, v_r, \Ss \overline{v_1},\ldots,\Ss\overline{v_r}, \Jj v_1,\ldots, \Jj v_r,\Jj \Ss \overline{v_1},\ldots, \Jj \Ss \overline{v_r}$.

\vspace{.1cm}

Indeed, if ${\rm dim}(\VM_0) =0$, there is nothing to prove. Otherwise
let $a_1^2 > 1$ be the biggest eigenvalue of $\Mm^*\Mm$ which is also the biggest eigenvalue of $\Mm^*\Mm$ restricted
to $\VM_0$ and let $v \in \VM_0$ be some corresponding eigenvector.
Then $\Jj \Ss \overline{v}$ is another eigenvector for the same eigenvalue.
Take $w=v+\mu \Jj \Ss \overline{v}$, where $\mu\in\CC$ can be chosen in such a way that
$w$ and $\Jj \Ss \overline{w}$ are orthogonal. Then also $\Jj w$ and $\Ss \overline{w}$ which are eigenvectors to the eigenvalue
$a_1^{-1}$ are orthogonal. As $a_1 > a_1^{-1}$, the space spanned by $w$ and $\Jj\Ss \overline{w}$ is orthogonal to the space
spanned by $\Jj W$ and $\Ss \overline{w}$. Therefore normalizing $w$ to $v_1 = w / \|w\|$ the vectors
$v_1, \Ss \overline{v_1}, \Jj v_1, \Jj \Ss \overline{v_1}$ are orthonormal. Denote the space spanned by these vectors
by $\VM_{0,1}\subset \VM_0$ and its orthogonal complement in $\VM_0$ by $\VM_{0,2}$ which is again
a $\GM$-like, $\Mm$-invariant subspace. One proceeds by induction to complete the proof of the claim.

\vspace{.1cm}

\noindent {\bf Second claim:} If $L=2d$, then ${\rm dim}(\VM_1)$ is divisible by 4 and there is an orthonormal basis of the form
$v_{r+1}, \ldots, v_d, \Ss \overline{v_{r+1}},\ldots,\Ss\overline{v_d}, \Jj v_{r+1},\ldots, \Jj v_d,\Jj \Ss \overline{v_{r+1}},\ldots, \Jj \Ss \overline{v_d}$.
If $L=2d+1$, then ${\rm dim}(\VM_1)$ is congruent to $2\!\! \mod 4$ and one has an orthonormal basis which is of the form

\noindent $v_{r+1}, \ldots, v_d,v_{d+1}, \Ss \overline{v_{r+1}},\ldots,\Ss\overline{v_d}, \Jj v_{r+1},\ldots, \Jj v_d, \Jj v_{d+1},
\Jj \Ss \overline{v_{r+1}},\ldots, \Jj \Ss \overline{v_d}$ with $\Ss v_{d+1} = \overline{v_{d+1}}$.

\vspace{.1cm}

Indeed, as $\Jj$ is unitary and operates on $\VM_1$, there is an orthonormal basis of $\VM_1$ of eigenvectors of $\Jj$.
The eigenvalues of $\Jj$ are $\pm \imath$. If $\Jj v = \pm\imath v$, then $\Jj \Ss \overline{v} = \Ss \Jj \overline{v} = \mp \imath \Ss \overline{v}$.
Hence the dimensions of the eigenspaces of $\Jj$ in $\VM_1$ are equal.
If ${\rm dim}(\VM_1) \geq 4$, there are two orthonormal vectors $w_1, w_2$ satisfying $\Jj w_j = \imath w_j$.
As $\Jj \Ss \overline{w_j} = - \imath \Ss \overline{w_j}$ the vectors $w_1, w_2, \Ss \overline{w_1}, \Ss \overline{w_2}$ are orthonormal.
Set $v_{r+1}=\frac{1}{\sqrt{2}} (w_1+\Ss\overline{w_2})$. Then the vectors $v_{r+1}$,
$\Jj v_{r+1} = \frac{\imath}{\sqrt{2}}(w_1-S\overline{w_2})$, $\Ss \overline{v_{r+1}}=\frac1{\sqrt{2}}(w_2+\Ss \overline{w_1})$ and
$\Jj\Ss \overline{v_{r+1}}=\frac1{\sqrt{2}} (w_2 -\Ss \overline{w_1})$ are orthonormal. They span a 4-dimensional
$\GM$-like subspace of $\VM_1$. Denote its orthonormal complement in $\VM_1$ by $\VM_2$ and proceed by induction
to obtain the vectors $v_{r+2}, \ldots, v_d$. In case $L=2d$ this shows the above claim; if $L= 2d+1$, one is left
with some 2-dimensional, $\GM$-like subspace $\VM_{d-r+1}$. This space is spanned by the orthonormal vectors
$w$ and $\Ss \overline{w}$ where $\Jj w = \imath w$. Set $v_{d+1} = \frac1{\sqrt{2}}(w+\Ss \overline{w})$,
then $v_{d+1}$ and $\Jj v_{d+1}$ form an orthonormal basis of $\VM_{d-r+1}$ and $\Ss v_{d+1}=\overline{v_{d+1}}$.

\vspace{.1cm}

\noindent {\bf Construction of $\Uu$:} From the first two steps we obtain an orthonormal basis of eigenvectors of $ \Mm^* \Mm$ of the form
$(v_1,\ldots,v_d,v_{d+1},\Ss \overline{v_1},\ldots,\Ss \overline{v_d},\Jj v_1,\ldots,\Jj v_{d+1},\Jj \Ss \overline{v_1},\ldots,\Jj\Ss \overline{v_d})$
in case $L=2d+1$ and the same without the entries containing $v_{d+1}$ if $L= 2d$.
The corresponding eigenvalues of $v_1,\ldots, v_d$ shall be denoted by
$a_1^2 \geq a_2^2 \geq \ldots a_d^2 \geq 1$. The eigenvalue corresponding to $v_{d+1}$ if $L= 2d+1$ is 1.
Denote the canonical basis of $\CC^{2L}$ by $e_i,\; i=1,\ldots,2L$.
Let us define the unitary matrix $\Uu$ by
$$
\begin{array}{lcllllcll}
L=2d  & & & & \qquad & L=2d+1 & \\
\Uu v_i &=&  e_i & i=1,\ldots,d & & \Uu v_i &=& e_i & i=1,\ldots,d+1 \\
\Uu\Ss \overline{v_i} & =& e_{i+d} &  i=1,\ldots,d & & \Uu\Ss \overline{v_i} &=& e_{i+d+1} & i=1,\ldots,d \\
\Uu\Jj v_i &=& - e_{i+2d} &i=1,\ldots,d & & \Uu \Jj v_i &=& -e_{i+2d+1} & i=1,\ldots,d+1 \\
\Uu \Jj \Ss \overline{v_i} &=& - e_{i+3d} & i =1,\ldots, d & & \Uu \Jj \Ss \overline{v_i} &=& - e_{i+3d+2} & i=1,\ldots,d\;.
\end{array}
$$
Then defining the diagonal matrix $\Dd$ as in the statement of the proposition, one has $\Uu  \Mm \Uu^*=\Dd$.
For $i=1,\ldots,d$, one has
$$
\begin{array}{l}
(\Uu^* \Jj \Uu)v_i\;=\; \Uu^* \Jj e_i = -\Uu^* e_{i+L} = \Jj v_i\;, \\
(\Uu^* \Jj \Uu)\Jj v_i \;=\; - \Uu^* \Jj e_{i+L} = - \Uu^*e_i = -v_i = \Jj (\Jj v_i)\;,
\end{array}
$$
similar calculations hold for $\Ss \overline{v_i}$, $\Jj \Ss \overline{v_i}$ and also $v_{d+1}, \Jj v_{d+1}$ in the case $L=2d+1$.
Thus one obtains $\Uu^* \Jj \Uu = \Jj$. It is a matter of calculation
to verify that $\Uu^t \Ss \Uu = \Ss$ and hence $\Uu \in \GM \cap {\rm U}(2L)$.
Finally, as $\Uu \in \GM$ we have $\Aa \Uu \Aa^* \in \SO^*(2L) \cap {\rm U}(2L) =\,$SP$(2L,\RR) \cap {\rm O}(2L)$
and hence $\det (\Uu) = \det(\Aa \Uu \Aa^*) = 1$ and therefore $\Uu \in {\rm SU}(2L)$. \\[.2cm]
(v) As $\Mm^*\Mm \in \GM$ and $\Mm^*\Mm>0 $, by (iv) we find $\Uu \in \GM \cap {\rm SU}(2L)$
and a diagonal matrix $\Dd$ as above, such that $\Uu \Mm^*\Mm \Uu^* = \Dd^2$.
Set $\Kk= \Mm \Uu^* \Dd^{-1} \in \GM$, then $\Mm=\Kk\Dd\Uu$ and $\Kk^*\Kk=\Dd^{-1} \Uu\Mm^* \Mm \Uu^* \Dd^{-1} = \one$.
Hence $\Kk\in \GM \cap {\rm U}(2L)=\GM \cap {\rm SU}(2L)$.\\[.2cm]
(vi) By (v), $\det(\Mm)=\det(\Kk) \det(\Dd) \det(\Uu)=1$.
Furthermore as the group SP$(2L,\RR)\cap{\rm O}(2L)$ is connected,
also $\SO^*(2L) \cap {\rm U}(2L)$ is.
Using the decomposition in (iv) one easily obtains that $\GM$ is connected.
\hfill $\Box$

\vspace{.2cm}

Now let $(\Yy_n)_{n\geq 1}$ be an i.i.d. sequence in $\GM$. Then by Lemma~\ref{lemma-decomp}
the whole associated Lyapunov spectrum has at least multiplicity two. So let
$\gamma_1,\gamma_1,\gamma_2,\gamma_2,\ldots,\gamma_L,\gamma_L$ be the $2L$ Lyapunov exponents
with $\gamma_1 \geq \gamma_2 \geq \ldots \gamma_L$.
Lemma~\ref{lemma-decomp} also shows $\gamma_p = - \gamma_{L+1-p}$ and in the case
$L=2d+1$, one has $\gamma_{d+1}= 0$. Therefore it is always enough to consider
$\gamma_1,\ldots, \gamma_d$.
Set $v^{(p)} = e_1 \wedge \ldots \wedge e_p \wedge e_{2L-d+1} \wedge \ldots e_{2L-d+p}$
and define $\LM_p={\rm span_\RR} (\{ \Lambda^{2p}\Mm v^{(p)}\,|\, \Mm \in \GM \})$ which is a real linear subspace
of $\Lambda^{2p} \CC^{2L}$. Note that $\LM_p$ does not have to be a complex vector space.
Taking the real part of the scalar product on $\Lambda^{2p} \CC^{2L}$ induces a scalar product on
$\LM_p$ but actually one does not need to take the real part as the following lemma shows.

\begin{lemma}
\label{lemma-scalar}
The scalar product in $\Lambda^{2p} \CC^{2L}$ of two vectors in $\LM_p$ is real.
Let $f_1, f_2,f_3, f_4 \in \LM_p$ and consider $f_1 \wedge f_2, f_3 \wedge f_4$
on one hand as elements in $\Lambda^2(\Lambda^{2p} \CC^{2L})$ and on the other hand as elements
in $\Lambda^2_\RR \LM_p$ considered as tensor product over the field $\RR$.
Then the scalar products coincide, {\rm i.e.}
$\langle f_1 \wedge f_2, f_3 \wedge f_4 \rangle_{\Lambda^2(\Lambda^{2p} \CC^{2L})}
\;=\; \langle f_1 \wedge f_2, f_3 \wedge f_4 \rangle_{\Lambda^2_\RR \LM_p}$.
\end{lemma}

\noindent {\bf Proof.}
One finds $\Jj \Ss e_i = -e_{2L-d+i}$ and $\Jj \Ss e_{2L-d+i}=e_i$ for $i=1,\ldots,d$ which implies
\mbox{$\Lambda^{2p} (\Jj \Ss) v^{(p)} = (-1)^{2p} v^{(p)} = v^{(p)}$}.
For $\Mm \in \GM$ one has $\Ss \Mm \Ss = \Jj^* \overline{\Mm} \Jj$ and hence
$$
\begin{array}{lcl}
\langle v^{(p)}, \Lambda^{2p} \Mm v^{(p)} \rangle
&=&
\langle \Lambda^{2p} \Ss v^{(p)}, \Lambda^{2p} (\Ss \Mm \Ss^2) v^{(p)} \rangle
\\
& = &
\langle \Lambda^{2p} (\Jj \Ss) v^{(p)}, \Lambda^{2p} (\overline{\Mm} \Jj \Ss) v^{(p)}\rangle 
\;=\;
\overline{\langle v^{(p)}, \Lambda^{2p} \Mm v^{(p)} \rangle}\;.
\end{array}
$$
Therefore $\langle \Lambda^{2p} \Mm v^{(p)}, \Lambda^{2p} \Nn v^{(p)}\rangle
= \langle v^{(p)}, \Lambda^{2p} (\Mm^* \Nn) v^{(p)}\rangle$ is real for all
$\Mm,\Nn \in \GM$ and by linearity the $\Lambda^{2p} \CC^{2L}$ scalar product for two vectors in $\LM_p$ is real. The second statement follows from the first one using
$\langle f_1 \wedge f_2, f_3 \wedge f_4 \rangle
= \langle f_1, f_3 \rangle \langle f_2, f_4 \rangle - \langle f_1, f_4 \rangle \langle f_2, f_3 \rangle$.
\hfill $\Box$

\vspace{.2cm}

Considering $f_1 \wedge f_2$ as element in
$\Lambda^2_\RR \LM_p$ on one hand and as an element of $\Lambda^2 (\Lambda^{2p}\CC^{2L})$ on the other
hand induces an $\RR$-linear map $\Lambda^2_\RR \LM_p \to \Lambda^2(\Lambda^{2p}\CC^{2L})$.
By Lemma~\ref{lemma-scalar} this map preserves the inner product and is hence injective.
Therefore $\Lambda^2_\RR \LM_p$ can be viewed as real subspace of
$\Lambda^2(\Lambda^{2p}\CC^{2L})$. The following criterion for
distinctness of the Lyapunov exponents is adapted from \cite{GR,BL}.

\begin{defini}
A subset $\TM$ of $\GM$ is $\LM_p$-strongly irreducible if there does not exist a finite union $\WM$ of
proper linear subspaces of $\LM_p$ such that $(\Lambda^{2p} \Mm)(\WM)=\WM$ for any $\Mm$ in $\TM$.
\end{defini}

\begin{proposi}\label{prop-dist-lyap1}
Let $(\Yy_n)_{n\geq 1}$ be a sequence of i.i.d. random matrices in $\GM$ for $L=2d$ or $L=2d+1$
and let $p$ be an integer $1 \leq p \leq d$. Let $\TM$ be the semi-group generated by the support of $\Yy_n$. Suppose that $\TM$ is $2p$-contracting and $\LM_p$-strongly irreducible and that
$\EE(\log_+\|\Yy_1\|)<\infty$. Then $\gamma_p > \gamma_{p+1}$.
\end{proposi}

\noindent {\bf Proof.}
Let $k$ be the dimension of $\LM_p$ and $(f_1,\ldots,f_k)$ an orthonormal basis to be chosen later on.
For any $\Mm \in \GM$ let $\widehat \Mm$ denote the matrix in ${\rm Gl}(k,\RR)$ with the entries
$$
\widehat \Mm_{i,j} \;=\;
\langle f_i, \Lambda^{2p} \Mm f_j \rangle\;, \qquad 1 \leq i, j \leq k\;.
$$
If $\Uu\in \GM \cap {\rm U}(2L)$, then
$\Lambda^{2p}\Uu \in \Lambda^{2p} \GM \cap {\rm U}(\Lambda^{2p}\CC^{2L})$ and hence the restriction
of $\Lambda^{2p} \Uu$ to $\LM_p$ is orthogonal, {\it i.e. } $\widehat \Uu \in {\rm O}(\LM_p)$.
Let us use the notation $\Lambda^{2p}\Mm = \Phi(\Mm)$ .
One has $\|\widehat \Mm\| \leq \|\Phi(\Mm)\|$ as $\LM_p$ is a subspace of $\Lambda^{2p} \CC^{2L}$
and by Lemma~\ref{lemma-scalar} one also obtains $\|\Lambda^2 \widehat \Mm\|\leq\|\Lambda^2 \Phi(\Mm)\|$.

\vspace{.1cm}

\noindent {\bf Claim:} Let $a_1 \geq a_2 \geq \ldots \geq a_d \geq 1$ be
the singular values of $\Mm$ as occurring in the decomposition in Lemma~\ref{lemma-decomp}(v), then
$\|\Phi(\Mm)\| = a_1^2 \cdots a_p^2 = \|\widehat \Mm\|$ and
$\|\Lambda^2 \Phi(\Mm)\| \geq \|\Lambda^2_\RR \widehat \Mm \|
\geq \| \widehat \Mm \| \cdot a_1^2 \cdots a_{p-1}^2 a_{p+1}^2$. In the case $p=d$, we define $a_{d+1}=a_d^{-1}$.

\vspace{.1cm}

Indeed, set $f_1 =v^{(p)}=e_1 \wedge \ldots \wedge e_p \wedge e_{2L-d+1} \wedge \ldots \wedge e_{2L-d+p}$
and if $p< d$ set $f_2 = e_1 \wedge \ldots \wedge e_{p-1} \wedge e_{p+1} \wedge e_{2L-d+1} \wedge \ldots
\wedge e_{2L-d+p-1} \wedge e_{2L-d+p+1}$.
In the case $p=d$, set
$f_2 = e_1 \wedge \ldots \wedge e_{d-1} \wedge e_{L+d} \wedge e_{2L-d+1} \wedge \ldots
\wedge e_{2L-1} \wedge e_{L}$. Further, for any $d \times d$ invertible matrix $B$ and any
matrix $C$ with $B^* C = C^* B$, one can construct
the following element of $\GM$:
\begin{equation} \label{eq-prototype}
\Nn\;=\;
\left(\begin{matrix}
B & 0 & 0 & 0 & 0 & 0 \\ 0 & \cos(\varphi) & 0 & 0 & \sin(\varphi) & 0 \\
0 & 0 &(B^t)^{-1} & 0 & 0 & -\overline{C} \\
{C} & 0 & 0 & (B^*)^{-1} & 0 & 0 \\ 0 & -\sin(\varphi) & 0 & 0 & \cos(\varphi) & 0 \\
0 & 0 & 0 & 0 & 0 & \overline{B}
\end{matrix} \right)\;,
\qquad \begin{array}{l}
\text{if $L=2d$ pencil out} \\
\text{the rows and columns} \\
\text{containing $\varphi$.}
\end{array}
\end{equation}
Thus for $p<d$, one readily finds $\Nn\in \GM$ with $f_2= \Lambda^{2p} \Nn f_1 \in \LM_p$.
In the case $p=d$ define $\Nn_1$ by setting $B=\one$ and $C_{i,j}=0$ except $C_{d,d}=1$ and
define $\Nn_2$ by setting $B=2\cdot \one,\; C=0$.
Then one obtains
$(2^{2(d-1)}-2^{2(d-2)})f_2= (2^{2(d-1)}-\Lambda^{2p} \Nn_2)(\Lambda^{2p} \Nn_1 f_1 - f_1)\in L_d$.
In conclusion, $f_1, f_2 \in \LM_p$ can be completed to an orthonormal basis of $\LM_p$.
Now let us write $\Mm=\Kk\Dd\Uu$ as in Lemma~\ref{lemma-decomp}(v), then
$$
\|\Phi(\Mm)\| \;=\; a_1^2 \cdots a_p^2 \;= \;\|\Lambda^{2p} \Dd f_1\| \;=\; \|\widehat \Dd f_1\|
\;\leq \;\|\widehat \Dd\| \;\leq \;\|\Lambda^{2p} \Dd\|\;=\;\|\Phi(\Mm)\|
$$
where the last inequality holds as $\LM_p$ is a subspace of $\Lambda^{2p}\CC^{2L}$.
Hence $\|\Phi(\Mm)\|=\|\widehat \Dd\|$, but $\|\widehat \Dd\|=\|\widehat \Kk \widehat \Dd \widehat \Uu\|=\|\widehat \Mm\|$.
As mentioned above,  $\|\Lambda^2 \Phi(\Mm)\| \geq \| \Lambda^2 \widehat \Mm \| $.
Furthermore one has
$$
\|\Lambda^2 \widehat \Mm\| \;=\; \|\Lambda^2 \widehat \Dd \|
\;\geq\; \|\Lambda^2 \widehat \Dd (f_1\wedge f_2)\| \;=\;
\|\widehat \Mm\| \cdot a_1^2 \cdots a_{p-1}^2 a_{p+1}^2
$$
Hence the claim is proved.

\vspace{.1cm}

Let $\widehat \TM$ be the semi group induced by the distribution of $\widehat \Yy_1$.
As $\TM$ is $\LM_p$-strongly irreducible, clearly $\widehat \TM$ is a strongly irreducible subset
of ${\rm Gl}(k,\RR)$.
As $\TM$ is also $2p$ contracting, there exists a sequence $(\Mm_n)_{n\geq 1}$ in $\TM$ such that
$\lim_{n\to\infty} \|\Phi(\Mm_n)\|^2 \|\Lambda^2 \Phi(\Mm_n)\|^{-1}=\infty$.
As $\|\widehat \Mm_n\|=\|\Phi(\Mm_n)\|$ and $\|\Lambda^2 \Phi(\Mm_n)\|\geq\|\Lambda^2 \widehat \Mm_n \|$ by the above claim, one obtains
$$
\lim_{n\to\infty} \|\widehat \Mm_n\|^2 \|\Lambda^2 \widehat \Mm_n\|^{-1}
\;\geq\;
\lim_{n\to\infty} \|\Phi(\Mm_n)\|^2 \|\Lambda^2 \Phi(\Mm_n)\|^{-1}
\;=\;
\infty\;.
$$
Hence $\widehat \TM$ is contracting.
The two biggest Lyapunov exponents associated to the sequence $(\widehat \Yy_n)_{n\geq 1}$
shall be denoted by $\hat \gamma_1$ and $\hat \gamma_2$.
Then by the claim, the definition of Lyapunov exponents and
\cite[A.III.6.1]{BL}
one has
$$
2 \sum_{i=1}^p \gamma_i
\;=\;
\hat \gamma_1
\; > \;
\hat \gamma_2
\;\geq\;
2 \sum_{i=1}^{p-1} \gamma_i\;+\; 2 \gamma_{p+1}\;,
$$
implying $\gamma_p > \gamma_{p+1}$.
By definition of $a_{p+1}$ one actually would have to replace $\gamma_{p+1}$ by $\gamma_{p+2}=\gamma_{d+2}$ in the
case $L=2d+1, p=d$. Then one gets $\gamma_d > \gamma_{d+2}=-\gamma_d$ and therefore
$\gamma_d > 0 = \gamma_{d+1}$.
\hfill $\Box$

\vspace{.2cm}

\begin{theo}\label{theo-dist-lyap2}
Let $(\Yy_n)_{n\geq 1}$ be a sequence of i.i.d. random matrices in $\GM$ for $L=2d$ or $L=2d+1$
Let $\TM$ be the semi-group induced by the support of $\Yy_1$ and let $\EE(\log_+\|\Yy_1\|)<\infty$.
Suppose that $\TM$ is Zariski dense in $\GM$, then all Lyapunov exponents are distinct.
\end{theo}

\noindent {\bf Proof.}
According to the proof of Proposition~\ref{prop-dist-lyap1} the inequality $\gamma_p > \gamma_{p+1}$ follows
from the fact that the semi-group $\widehat \TM = \{\widehat \Mm \,|\, \Mm \in \TM\}$ is strongly irreducible and contracting
in ${\rm GL}(k,\RR)$ as defined above.
Now $\widehat \TM$ is Zariski dense in $\widehat \GM = \{\widehat \Mm\,|\, \Mm \in \GM \}$.
Otherwise there would be a polynomial
$\widehat{P}$ on ${\rm GL}(k,\RR)$ such that $\widehat{P}(\widehat \TM)=0$ and $\widehat{P}(\widehat \Mm) \neq 0$ for some $\Mm \in \GM$. As the entries in $\widehat \Mm$ are polynomials
of the entries in $\Mm$, this leads to a polynomial $P$ on ${\rm GL}(2L,\CC)$ such that $P(\TM)=0$ and
$P(\Mm) \neq 0$ for some $\Mm\in \GM$, contradicting the fact that $\TM$ is Zariski dense in $\GM$.

\vspace{.2cm}

Now suppose $\widehat \TM$ is not strongly irreducible. Then there would be a finite union of proper
subspaces $\WM= \VM_1 \cup \ldots \cup \VM_n$ such that $\Mm(\WM) \subset \WM$ for all $\Mm \in \widehat \TM$.
The property $\Mm(\VM_i) \subset \VM_k$ can be written as
$\langle w, \Mm v \rangle = 0$ for all $w \in \VM_k^\perp, v \in \VM_i$. Hence the set of all such matrices $\Mm$ is
Zariski closed. The property $\Mm(\WM) \subset \WM$ is therefore a finite intersection of finite unions of
Zariski closed sets and hence Zariski closed. As $\widehat \TM$ is Zariski dense in $\widehat \GM$, this then implies
$\widehat \GM (\WM) \subset \WM$. Therefore, if $\widehat \GM$ is strongly irreducible, then also $\widehat \TM$ is.

\vspace{.1cm}

To show that $\widehat \TM$ is contracting we want to use Theorem~6.3 of \cite{MG} which states that if the algebraic closure
of $\widehat \TM$ is strongly irreducible and contracting, then also $\widehat \TM$ is contracting.
Hence it is only left to show that $\widehat \GM$ is strongly irreducible and contracting.

\vspace{.1cm}

The property of $\widehat \GM$ to be strongly irreducible is equivalent to
$\GM$ being $\LM_p$-strongly irreducible.
As $\GM$ is connected we have to show that there is no proper subspace  $\VM\subset \LM_p$ such that
$(\Lambda^{2p} \Mm)(\VM)\subset\VM$ for all $\Mm\in\GM$.
Suppose such a $\VM$ exists.
For $a_1 > a_2 > \ldots >a_d > 1$ take $\Dd=\diag(a_1,\ldots, a_d,1,a_1^{-1},\ldots,a_d^{-1},a_1^{-1},\ldots,a_d^{-1},1,a_1,\ldots,a_d)$. The relation
$(\Lambda^{2p} \Dd^n )(\VM)\subset \VM$ implies that either $v^{(p)} \in \VM$, but then
$\LM_p=\VM$ or that $v^{(p)}$ is in the orthogonal complement $\VM^\perp$.
But then by Lemma~\ref{lemma-decomp}(i) one has,  for $v \in \VM$ and
any $\Mm\in\GM$,
$\langle \Lambda^{2p} \Mm v^{(p)}, v \rangle=
\langle v^{(p)},\Lambda^{2p} \Mm^* v \rangle =0
$. Hence $\LM_p = \VM^\perp$. Therefore $\VM$ is not proper.

\vspace{.1cm}

Now it is only left to show that $\widehat \GM$ is contracting.
By the proof of Proposition~\ref{prop-dist-lyap1} this follows if $\GM$ is $2p$-contracting.
Therefore take a matrix $\Mm$ of the form \eqref{eq-prototype} with
$C=0$ and $B=\diag(\lambda_1,\ldots,\lambda_d)$. such that all moduli of the eigenvalues are distinct
except for the fact that always two eigenvalues have the same modulus.
The sequence $\Mm^n$ then shows that $\GM$ is $2p$-contracting.
\hfill $\Box$.

\vspace{.2cm}

\noindent {\bf Proof of Theorem~\ref{theo-main}(i) and first claim of (ii).} The
Coupling Hypothesis implies by Theorem~\ref{theo-dist-lyap2} that the
Lyapunov exponents as defined in Section~\ref{sec-TRI} are distinct
apart from Kramers' degeneracy. The symplectic
symmetry of the Lyapunov spectrum implies that no
Lyapunov exponent vanishes for even $L$, while for odd $L$ there are exactly two
vanishing Lyapunov exponents. 
By Theorem~\ref{theo-Kotani} the absolutely continuous
spectrum is absent for even $L$ and has multiplicity 2 for odd $L$.
\hfill $\Box$.

\section{Absence of singular spectrum}
\label{sec-Absence}

In this section we only consider the random model described at the end of Section~\ref{sec-random}.
For any configuration $\omega=\left((\lambda_{j,k})_{k=1,\ldots,K ; j \in \ZZ}, (\Vv_j)_{j\in\ZZ},s\right)\in \Omega$ let
$\tilde\omega$ denote $\omega$ excluded the singular potential $\Vv=\Vv_0$ at $s$, {\it i.e.}
$\tilde \omega=\left((\lambda_{j,k})_{k=1,\ldots,K ,\, j \in \ZZ}\,,\, (\Vv_j)_{j\in\ZZ,j\neq 0},s\right)$.
The distribution of $\tilde \omega$ shall be denoted by $\tilde \PP$ and that of $\Vv$ by $\pp_\Vv$.
With these notations $\PP=\tilde \PP \times \pp_\Vv$.
We only consider the case where $L$ is odd and $\pp_\Vv$ is absolutely continuous
w.r.t. to the Lebesgue measure. Next recall the definition \eqref{eq-Vhat} of
$\widehat \Vv \in \Jj {\rm so}^*(2L)$.
Note that $\widehat \Vv$ is only defined for almost every $\Vv$ and for almost every
$\widehat \Vv$ there is a pre-image $\Vv$, which is not necessarily unique.
Furthermore the pre-images of zero sets are zero sets
and hence the distribution $\pp_{\widehat \Vv}$ of $\widehat \Vv$, {\it i.e.}
the image measure of $\pp_\Vv$, is absolutely continuous w.r.t. the Lebesgue
measure on the vector space $\Jj{\rm so}^*(2L)$.

\vspace{.1cm}

As $\Vv$ denotes the singular potential at $x_0=s$, let $\widehat G^z_\Vv$ denote the averaged Green matrix at the point $x_0=s$, that is,
$\widehat G^z_\Vv=\widehat G^z_\Vv(s)$ with the notations of Proposition~\ref{prop-GreenHerg}. Note that this matrix actually depends on $\omega=(\tilde \omega,\Vv)$, but in most of the arguments below $\tilde \omega$ will be fixed. Furthermore, Proposition~\ref{prop-GreenPerturb} shows that $\widehat G^z_\Vv$ actually only depends on $\widehat{\Vv}$ (which is a real statement statement since the map $\Vv\mapsto \widehat \Vv$ is not injective). Hence it is sufficient to prove almost sure statements w.r.t. the distribution $\pp_{\widehat \Vv}$ of $\widehat \Vv$ instead of w.r.t. the distribution $\pp_\Vv$ of $\Vv$.

\vspace{.1cm}

Let $\mu_\omega=\mu_{\tilde\omega,\Vv}$ denote the associated positive matrix valued measure.
The function $E\mapsto\frac{1}{1+E^2}$ is in $L^1(\mu_\omega)$ for all $\omega$.
On the set of such measures one may introduce the weak-$\ast$ topology induced by the functions
$E\mapsto \Im m((E-z)^{-1})$ for $z$ in the upper half plane.
As the pairing of this function with the measure $\mu_\omega$ is just $\Im m(G^z)$, it follows that the
map $\omega\mapsto \mu_\omega$ is Borelian. Finally let $\mu_{\omega,k}=\mu_{\tilde\omega,\Vv,k}$ denote the measure corresponding to $e_k^* \widehat G^z_\Vv e_k$ where $e_k$ is the $k$-th canonical basis vector of $\CC^{2L}$.

\vspace{.1cm}

The aim of this section is to prove that almost surely in $\omega$ the measure $\mu_\omega$ is absolutely continuous or equivalently, that its singular part vanishes, {\it i.e.} $\mu_{\omega,{\rm sing}}(\RR)=0$.
Therefore we will first show that almost surely one only needs to consider $\mu_{\omega,1}$ and then we show that $\mu_{\omega,1,{\rm sing}}(\RR)=0$ almost surely.
To obtain the first part we compare the measures $\mu_{\tilde\omega,\Vv,1}$ and $\mu_{\tilde\omega,\Vv,k}$
for fixed $\tilde\omega$ and show that they are almost surely equivalent.
Once cyclicity issues are settled (Proposition~\ref{prop-as-invert}) and matrix analogues of rank one perturbation results are proved (Proposition~\ref{prop-rank1like}), the proofs are basically modifications of the arguments of \cite{JL}.
Our starting point are the following observations linked to Kramers' degeneracy.

\begin{lemma} \label{lemma-start-id}
For $1\leq k,l \leq L$ let us introduce the $2L \times 2$ matrix $\Psi_k=(e_k,e_{k+L})$.

\vspace{.1cm}

\noindent {\rm (i)} Let $j$ denote the $2 \times 2$ symplectic form, then $\Jj \Psi_k = \Psi_k j$.

Furthermore one has $\Psi_k \Psi_k^* \in \Jj{\rm so}^*(2L)$ and
$\Psi_k j \Psi_l^* + \Psi_l j^* \Psi_k \in \Jj{\rm so}^*(2L)$.

\vspace{.1cm}

\noindent {\rm (ii) }
For $\Yy_1, \Yy_2 \in \Jj{\rm so}^*(2L)$ one has $\Yy_1 \Yy_2 \Yy_1 \in \Jj{\rm so}^*(2L)$.

\vspace{.1cm}

\noindent {\rm (iii)}  $\Psi_k^* \widehat G^z_\Vv \Psi_k$ is a multiple of the unity matrix, which means
 		$\Psi_k^* \widehat G^z_\Vv \Psi_k = e_k^* \widehat G^z_\Vv e_k\; \one$.

\end{lemma}

\noindent {\bf Proof.}
%
The identity $\Jj \Psi_k=\Psi_k j$ is readily verified. Furthermore
$(\Psi_k \Psi_k^*)^* = \Psi_k \Psi_k^*$ and one has
$\Jj^* \Psi_k \Psi_k^* \Jj= \Psi_k j^* j \Psi_k^* = \Psi_k \Psi_k^* = (\Psi_k \Psi_k^*)^t$
showing $\Psi_k \Psi_k^* \in \Jj{\rm so}^*(2L)$.
Similar calculations show $\Psi_k j \Psi_l^* + \Psi_l j^* \Psi_k \in \Jj{\rm so}^*(2L) \in \Jj{\rm so}^*(2L)$ and (i)
is proved. To obtain (ii), first note  that $\Yy_1,\Yy_2$ are self-adjoint and hence $\Yy_1\Yy_2\Yy_1$ is self-adjoint.
Furthermore one has
$\Jj^* \Yy_1\Yy_2\Yy_1 \Jj=\Jj^* \Yy_1 \Jj \Jj^* \Yy_2 \Jj \Jj^* \Yy_1 \Jj
=\Yy_1^t \Yy_2^t \Yy_1^t = (\Yy_1 \Yy_2 \Yy_1)^t$
and  also (ii) is proved.
(iii) is just a special case of Proposition~\ref{prop-GreenSymmetry}. \hfill $\Box$

\vspace{.2cm}

The measure class of $\mu_\omega$ is given by the trace, {\it i.e.} by the sum
$\sum_{k=1}^{2L} \mu_{\omega,k} = 2 \sum_{k=1}^L \mu_{\omega,k}$, where
the last identity follows from Lemma~\ref{lemma-start-id}(iii).

\begin{proposi} \label{prop-as-invert}
For fixed $\tilde\omega $, one has that for Lebesgue almost all $\widehat \Vv \in \Jj {\rm so}^*(2L)$
the set of energies $\{E \in \RR\,|\, \widehat G_\Vv^{E+\imath0} \;\text{\rm exists and}\; \Psi_l^* \widehat G^E_\Vv \Psi_k \;
\text{\rm is invertible}\,\}$ has full Lebesgue measure.
\end{proposi}

\noindent {\bf Proof}. We first claim that for fixed $z$ in the upper half plane $\UM_1$, there is a $\widehat{\Vv} \in \Jj {\rm so}^*(2L)$ such that $\Psi_l^* \widehat G^z_\Vv \Psi_k$ is invertible.
Recall that $\widehat G^z_\Vv=((\widehat G^z_0)^{-1}+\widehat \Vv)^{-1}$.
Set $(\widehat G^z_0)^{-1}=\Xx-\imath\,\Yy^{-1}$ with $\Yy^{-1} =-\Im m((\widehat G^z_0)^{-1})>0$.
As $\Jj^* \widehat G^z_0 \Jj = (\widehat G^z_0)^t$, one has $\Xx,\Yy^{-1},\Yy \in \Jj {\rm so}^*(2L)$.
Then consider $\widehat \Vv = -\Re e((\widehat G^z_0)^{-1}) + \lambda {\cal P}$ with a perturbation ${\cal P}\in \Jj {\rm so}^*(2L)$. Then
$$
\widehat G^z_\Vv\;=\;
(-\imath \Yy^{-1}+\lambda {\cal P})^{-1}\; =\; \imath\,\Yy+\lambda\, \Yy{\cal P}\Yy - \imath\, \lambda^2 \Yy{\cal P}\Yy{\cal P}\Yy + \Oo(\lambda^3)\;.
$$
Note that $\Vv$  now depends on $\lambda$ and ${\cal P}$, furthermore $\Yy{\cal P}\Yy \in \Jj {\rm so}^*(2L)$ as well as $\Yy{\cal P}\Yy{\cal P}\Yy \in \Jj {\rm so}^*(2L)$ by Lemma~\ref{lemma-start-id}.
For any $2\times 2$ matrices $A,B,C$ one has
$\det(A+\lambda B + \lambda^2 C)=\det(A)+\lambda \Tr(A(j^*Bj)^t) +
\lambda^2\left( \det(B)+\Tr(A(j^*Cj)^t)\right)+\Oo(\lambda^3)$.
Furthermore for $\Ww \in \Jj {\rm so}^*(2L)$, one has $(j^* \Psi_l^* \Ww\Psi_k j)^t=j^* \Psi_k^* \Ww^t \Psi_l j = \Psi_k^* \Jj^* \Ww^t \Jj \Psi_l = \Psi_k^* \Ww\Psi_l$.
Thus from the above
\begin{eqnarray}
\det(\Psi_l^* \widehat G^z_\Vv \Psi_k) & = &
\imath \;\det(\Psi_l^* \Yy \Psi_k)\;+\;\imath\, \lambda \;\Tr( \Psi_l^* \Yy \Psi_k \Psi_k^* \Yy{\cal P}\Yy \Psi_l)\;+ \nonumber
\\
& &
\label{eq-det-exp}
+\;\lambda^2\;\bigl( \det(\Psi_l^* \Yy{\cal P}\Yy \Psi_k)\,-\,\imath \;\Tr( \Psi_l^*\Yy\Psi_k \Psi_k^* \Yy{\cal P}\Yy{\cal P}\Yy \Psi_l) \bigr)
\;+\;\Oo(\lambda^3)\;.
\end{eqnarray}
If $\det (\Psi_l^* \Yy \Psi_k) \neq 0$, then the claim is true (just take $\lambda=0$).
If $\det (\Psi_l^* \Yy \Psi_k) =0$, but $\Psi_l^* \Yy \Psi_k \neq 0$, then set ${\cal P}= \Yy^{-1} \in \Jj {\rm so}^*(2L)$ and
\eqref{eq-det-exp} reduces to
$$
\det(\Psi_l^* \widehat G^z_\Vv \Psi_k)\;=\;
\imath\,\lambda \; \Tr((\Psi_l^* \Yy \Psi_k)^*(\Psi_l \Yy \Psi_k))\; +\; \Oo(\lambda^2)\;.
$$
Since the coefficient before $\lambda$ only vanishes if $\Psi_l^* \Yy \Psi_k =0$,
this is not equal to zero for small $\lambda$ and the claim holds again.
Finally, if $\Psi_l^* \Yy \Psi_k = 0$, then set ${\cal P}=\Yy^{-1} (\Psi_l j \Psi_k^* + \Psi_k j^* \Psi_l^*+\Psi_l\Psi_l^*) \Yy^{-1}$
which lies in $\Jj{\rm so}^*(2L)$ by Lemma~\ref{lemma-start-id}  part (i) and (ii).
Then \eqref{eq-det-exp} reduces to
$$
\det(\Psi_l^* \Yy \Psi_k) = \lambda^2 \det(\Psi_l^*\Psi_l j \Psi_k^*\Psi_k+\Psi_l^*(\Psi_k j^*+\Psi_l) \Psi_l^*\Psi_k) + \Oo(\lambda^3)
= \lambda^2 + \Oo(\lambda^3)\;,
$$
where we used $\Psi_l^* \Psi_k = \delta_{l,k}$.
Hence this determinant is again not zero for small $\lambda$.
Thus for all cases we find some $\widehat \Vv$ such that $\Psi_l^* \widehat G^z_\Vv \Psi_k$ is invertible and the claim is proved.

\vspace{.1cm}

Now by definition of the determinant and Cramer's rule the function
$\widehat \Vv \mapsto \det(\Psi_l^* \widehat G^z_\Vv \Psi_k)=\det(\Psi_l^* ((\widehat G^z_0)^{-1}+\widehat{\Vv})^{-1}\Psi_k)$
is a rational function on the vector space $\Jj {\rm so}^*(2L)$ which does not vanish completely by the claim above, therefore it does not vanish for
Lebesgue almost every $\widehat \Vv \in \Jj {\rm so}^*(2L)$ w.r.t. the Lebesgue measure on $\Jj {\rm so}^*(2L)$.

\vspace{.1cm}

Next recall that the boundary values $\widehat G_\Vv^{E+\imath 0} $ exist almost surely in $E$ by analyticity.
For $\widehat \Vv$ as described above, the map $z \mapsto \det(\Psi_l^* \widehat G^z_\Vv \Psi_k)$ is analytic in the upper half plane and does not vanish identically.
Therefore for Lebesgue almost every $E$, $\widehat G_\Vv^{E+\imath 0} $ exists and one has
$\det (\Psi_l^* \widehat G^{E+\imath0}_\Vv \Psi_k) \neq 0$.
\hfill $\Box$

\vspace{.1cm}

\begin{proposi}\label{prop-rank1like}
Let $\tilde \omega$ and $\widehat{\Vv} \in \Jj {\rm so}^*(2L)$ be fixed and define
$\widehat \Vv_\lambda = \widehat \Vv + \lambda \Psi_k \Psi_k^*$.

\vspace{.1cm}

\noindent {\rm (i)}
The set $ A_{\Vv_\lambda,k}=\{E\in\RR \,|\, \Psi_k^* \widehat G^{E+\imath0}_{\Vv_\lambda} \Psi_k \;\text{\rm exists and}\;
\Im m(\Psi_k^* \widehat G^{E+\imath0}_\Vv \Psi_k)>0\}$ is independent of $\lambda$

and  it is an essential support of the absolutely
continuous part of $\mu_{\tilde\omega,\Vv_\lambda,k}$.

\vspace{.1cm}

\noindent {\rm (ii)} The singular part of $\mu_{\tilde\omega,\Vv_\lambda,k}$ is supported on the set
$\{E\in\RR\,|\,\Psi_k^* \widehat  G^{E+\imath0}_{\Vv_0} \Psi_k = - \lambda^{-1}\,\one \}.$

\vspace{.1cm}

\noindent {\rm (iii)} For any $B\subset \RR$ of zero Lebesgue measure, we have
$\mu_{\tilde\omega\Vv_\lambda,k}(B)=0$ for Lebesgue a.e. $\lambda \in \RR$.

\end{proposi}

\noindent {\bf Proof.}
(i) We prove that $A_{\Vv,k}=A_{\Vv_0,k} \subset A_{\Vv_\lambda,k}$ for all $\lambda$; the other inclusion can be obtained analogously. Hence let $E \in A_{\Vv,k}$.
We first claim that $\one+\lambda \Psi_k \Psi_k^* \widehat G^{E+\imath 0}_{\Vv} $ is invertible.
Suppose $(\one+\lambda \Psi_k \Psi_k^* \widehat G^{E+\imath 0}_{\Vv} )v=0$. Then $v$ is in the range of $\Psi_k$ and there are $\alpha,\eta\in\CC$ such that $v= \alpha e_k + \beta e_{k+L}$.
We use $e_k^* \widehat G^{E+\imath0}_{\Vv} e_{L+k}=0=e_{L+k}^* \widehat G^{E+\imath0}_{\Vv} e_k$ following from
$\Jj^*\widehat G^{z}_{\Vv}\Jj=(\widehat G^{z}_{\Vv})^t$. Thus
$\alpha = - \lambda \alpha e_k^* \widehat G^{E+\imath0}_{\Vv} e_k$ and
$\beta =- \lambda \beta e_{k+L}^* \widehat G^{E+\imath0}_{\Vv} e_{k+L}$.
But as $\Im m(e_k^* \widehat G^{E+\imath0}_{\Vv} e_k) = \Im m(e_{k+L}^* \widehat G^{E+\imath0}_{\Vv} e_{k+L})>0$ for $E\in A_{\Vv,k}$
this implies $\alpha=0=\beta$ and hence $v=0$. Therefore the kernel of $\one+\lambda \Psi_k \Psi_k^* \widehat G^{E+\imath 0}_{\Vv} $ is indeed trivial.
Hence by Proposition~\ref{prop-GreenPerturb}, $\widehat G^{E+\imath0}_{\Vv_\lambda} =\widehat G^{E+\imath0}_{\Vv} (\one+\lambda \Psi_k \Psi_k^*\widehat G^{E+\imath 0}_{\Vv})^{-1}$
exists. Furthermore, also by Proposition~\ref{prop-GreenPerturb},
$$
\Im m(\widehat G^{E+\imath0}_{\Vv_\lambda} )\;=\;
\left[(\one+\lambda \Psi_k \Psi_k^*\widehat G^{E+\imath 0}_{\Vv_0})^{-1}\right]^*
\Im m(\widehat G^{E+\imath0}_{\Vv_0})
( \one+\lambda \Psi_k \Psi_k^*\widehat G^{E+\imath 0}_{\Vv_0})^{-1}\;,
$$
and $(\one+\lambda \Psi_k \Psi_k^*\widehat G^{E+\imath 0}_{\Vv_0})^{-1}$ leaves the space spanned by $e_{k}$ and $e_{k+L}$
invariant. Therefore one also obtains
$\Im m(\Psi_k^* \widehat G^{E+\imath0}_{\Vv_\lambda} \Psi_k)>0$ showing $E\in A_{\Vv_\lambda,k}$.

\vspace{.1cm}

(ii) From \eqref{eq-GreenPerturb1},
\begin{equation}
\label{eq-rank1pert1}
\widehat G^z_{\Vv_\lambda}\; = \;\widehat G^z_\Vv + \widehat G^z_\Vv[(\widehat G^z_\Vv)^{-1}-(\widehat G^z_{\Vv_\lambda})^{-1}] \widehat G^z_{\Vv_\lambda}
\;= \;\widehat G^z_{\Vv} - \lambda\, \widehat G^z_\Vv \Psi_k \Psi_k^* \widehat G^z_{\Vv_\lambda}\;,
\end{equation}
and hence $\Psi_k^* \widehat G^z_{\Vv_\lambda} \Psi_k = (1+\lambda \Psi_k^* \widehat G^z_\Vv \Psi_k)^{-1} \Psi_k^* \widehat G^z_\Vv \Psi_k$. Thus Lemma~\ref{lemma-start-id}(iii) implies
\begin{equation} \label{eq-rank1pert2}
e_k^* \widehat G^z_{\Vv_\lambda} e_k \,=\,
(1+\lambda\, e_k^* \widehat G^z_\Vv e_k)^{-1} \,e_k^* \widehat G^z_\Vv e_k\;.
\end{equation}
Thus in the limit $\epsilon \downarrow 0$,
$e_k^* \widehat G^{E+\imath\epsilon}_{\Vv_\lambda} e_k \to \infty$ if and only if
$\Psi_k^* \widehat G^{E+\imath \epsilon}_\Vv \Psi_k \to -\lambda^{-1}$.

\vspace{.1cm}

(iii) From \eqref{eq-rank1pert2} one deduces that the map $\lambda \mapsto \mu_{\tilde\omega,\Vv_\lambda,k}$ is
integrable in the *-weak topology over intervals $[a,b]$.
Taking imaginary parts of \eqref{eq-rank1pert2}, one obtains
$$
\Im m(e_k^* \widehat G^z_{\Vv_\lambda} e_k)\;=\;
\frac{\Im m(e_k^* \widehat G^z_\Vv e_k)}
{(1+\lambda \,\Re e(e_k^* \widehat G^z_\Vv e_k))^2+(\lambda \,\Im m(e_k^* \widehat G^z_\Vv e_k))^2} \;.
$$
Let $x=\Re e(e_k^* \widehat G^z_\Vv e_k)$ and $y=\Im m(e_k^* \widehat G^z_\Vv e_k)$. Then $\arctan(\frac{x^2+y^2}{y}\,\lambda+ \frac{x}{y})$ is an anti-derivative of the function $\lambda\mapsto\Im m(e_k^* \widehat G^z_{\Vv_\lambda} e_k)$.
Therefore $\int_{-a}^b d\lambda\,\Im m(e_k^* \widehat G^z_{\Vv_\lambda} e_k)$ is bounded by $\pi$
and the integral over the whole real line exists and is equal to $\pi$.
This means that the integral $\int_{-\infty}^\infty d\lambda \mu_{\tilde\omega,\Vv_\lambda,k}$ actually converges to the Lebesgue
measure which has no singular part.

\vspace{.1cm}

Now let $B$ be a set of Lebesgue measure zero. Then $\int_{-\infty}^\infty d\lambda \,\mu_{\tilde\omega,\Vv_\lambda,k}(B) = 0$.
As the measures are positive this means that for Lebesgue a.e. $\lambda \in \RR$ one has $\mu_{\tilde\omega,\Vv_\lambda,k}(B)=0$.
\hfill $\Box$

\vspace{.2cm}

Note that the equation proved in part (iii) above,
$dE=\int_\RR d\lambda\,\mu_{\tilde\omega,\Vv_\lambda,k}(dE)$,
is well-known from the theory of rank one perturbations.

\vspace{.1cm}

\begin{theo}
\label{theo-ac-relation}
Let $\omega=(\tilde \omega,\Vv)$ be fixed such
that the matrices $\Psi_1^* \widehat G^{E+\imath 0}_{\Vv} \Psi_{k}$, $\Psi_1^* \widehat G^{E+\imath 0}_{\Vv} \Psi_1$ as well as
$\Psi_k^* \widehat G^{E+\imath 0}_{\Vv} \Psi_k$ exist and are invertible for Lebesgue almost all $E$.
Set $\widehat \Vv_\lambda = \widehat \Vv + \lambda \Psi_k \Psi_k^*$.
Then for Lebesgue almost all $\lambda \in \RR$, the measure $\mu_{\tilde\omega,\Vv_\lambda,k}$ is absolutely continuous w.r.t. $\mu_{\tilde\omega,\Vv_\lambda,1}$.
\end{theo}

\noindent {\bf Proof.} By the Radon-Nikodym theorem we can decompose the measure $\mu_{\tilde\omega,\Vv_\lambda,k}=
f_{\lambda} \,\mu_{\tilde\omega,\Vv_\lambda,1} + \tilde\mu_\lambda$
where $f_\lambda$ is a function and $\tilde\mu_\lambda$ is the part of $\mu_{\tilde\omega,\Vv_\lambda,k}$ which is singular to $\mu_{\tilde\omega,\Vv_\lambda,1}$.
The statement of the theorem is that $\tilde\mu_\lambda=0$ for Lebesgue almost all $\lambda$.

\vspace{.1cm}

In order to show this, we first need to verify a few identities. By multiplying \eqref{eq-rank1pert1} with $\Psi_k^*$ from the left and $\Psi_1$ from the right, one obtains
\begin{equation} \label{eq-rank1pert3}
\Psi_k^* \widehat G^z_{\Vv_\lambda} \Psi_1\;=\;
(\one+\lambda \Psi_k^* \widehat G^z_\Vv \Psi_k)^{-1} \Psi_k^* \widehat G^z_\Vv \Psi_1
\;=\; \frac{\Psi_k^* \widehat G^z_\Vv \Psi_1}{1+\lambda\, e_k^* \widehat G^z_\Vv e_k}
\end{equation}
where the last identity follows from Lemma~\ref{lemma-start-id}(iii).
From \eqref{eq-rank1pert1}, one also obtains
\begin{equation} \label{eq-rank1pert4}
\Psi_1^* \widehat G^z_{\Vv_\lambda} \Psi_1 \;=\;\Psi_1^* \widehat G^z_{\Vv} \Psi_1
-\lambda \Psi_1^* \widehat G^z_{\Vv} \Psi_k \Psi_k^* \widehat G^z_{\Vv_\lambda} \Psi_1\;.
\end{equation}
Inserting \eqref{eq-rank1pert3} in \eqref{eq-rank1pert4} gives
\begin{equation} \label{eq-rank1pert5}
\Psi_1^* \widehat G^z_{\Vv_\lambda} \Psi_1 \;=\;\Psi_1^* \widehat G^z_{\Vv} \Psi_1
-\lambda\;\frac{\Psi_1^* \widehat G^z_\Vv \Psi_k \Psi_k^* \widehat G^z_\Vv \Psi_1}{1+\lambda\, e_k^* \widehat G^z_\Vv e_k}\;.
\end{equation}
Furthermore, it follows from \eqref{eq-rank1pert2} that
\begin{equation} \label{eq-rank1pert6}
1+\lambda\, e_k^* \widehat G^z_\Vv e_k\;=\; \frac{e_k^* \widehat G^z_{\Vv}e_k }{e_k^* \widehat G^z_{\Vv_\lambda} e_k}\;.
\end{equation}
\vspace{.1cm}

Now let $A \subset \RR$ be the set of all $E$ where the limit $\widehat G^{E+\imath0}_\Vv$ exists and all four matrices $\Psi_k^* \widehat G^{E+\imath0}_\Vv \Psi_k$, $\Psi_1^* \widehat G^{E+\imath0}_\Vv \Psi_1,
\Psi_1^* \widehat G^{E+\imath0}_\Vv \Psi_k$ and $\Psi_k^* \widehat G^{E+\imath0}_\Vv \Psi_1$ are invertible.
By assumption, the set $A$ has full Lebesgue measure and thus by Proposition~\ref{prop-rank1like}(iii)
we have $\mu_{\tilde\omega,\Vv_\lambda,k}=\mu_{\tilde\omega,\Vv_\lambda,k}|A$ for Lebesgue a.e. $\lambda \in \RR$. Thus we can restrict the measures to the set $A$. We consider the absolutely continuous and singular part of $\mu_{\tilde\omega,\Vv_\lambda,k}$ (w.r.t. the Lebesgue measure) separately and begin with the singular part. Inserting \eqref{eq-rank1pert6} into \eqref{eq-rank1pert5} and dividing by $e_k^* \widehat G^z_\Vv e_k$ gives
$$
\frac{\Psi_1^* \widehat G^z_{\Vv_\lambda}\Psi_1}{e_k^* \widehat G^z_{\Vv_\lambda} e_k}
\;=\;
\frac{\Psi_1^* \widehat G^z_{\Vv}\Psi_1}{e_k^* \widehat G^z_{\Vv_\lambda} e_k}
\,-\, \lambda \frac{\Psi_1^* \widehat G^z_\Vv \Psi_k \Psi_k^* \widehat G^z_\Vv \Psi_1 }{e_k^* \widehat G^z_\Vv e_k}\;.
$$
Let $E\in A$. Then taking $z=E+\imath \epsilon$ and the limit $\epsilon \downarrow 0$, it follows that
$$
\lim_{\epsilon\downarrow 0} \,\frac{\Psi_1^* \widehat G^{E+\imath \epsilon}_{\Vv_\lambda} \Psi_1}{e_k^* \widehat G^{E+\imath \epsilon}_{\Vv_\lambda} e_k}
\;=\;\lim_{\epsilon \downarrow 0} \frac{\Psi_1^* \widehat G^{E+\imath \epsilon}_{\Vv} \Psi_1}
{e_k^* \widehat G^{E+\imath \epsilon}_{\Vv_\lambda} e_k}\,-\,
\lambda\; \frac{\Psi_1^* \widehat G^{E+\imath 0}_\Vv \Psi_k \Psi_k^* \widehat G^{E+\imath0}_\Vv \Psi_1}
{e_k^* \widehat G^{E+\imath0}_{\Vv} e_k}\;.
$$
where the last term exists and is not zero (except for $\lambda=0$) by the invertibility assumptions for $E \in A$.
Since $|e_k^* \widehat G^{E+\imath \epsilon}_{\Vv_\lambda} e_k|\to \infty$ as $\epsilon \downarrow 0$ for a.e.
$E$ w.r.t. the singular part of $\mu_{\tilde\omega,\Vv_\lambda,k}$ and since, by Lemma~\ref{lemma-start-id}(iii), the matrix on the l.h.s. is a multiple of $\one$, one obtains
$$
\lim_{\epsilon \downarrow 0}
\frac{e_1^* \widehat G^{E+\imath \epsilon}_{\Vv_\lambda} e_1}{e_k^* \widehat G^{E+\imath \epsilon}_{\Vv_\lambda} e_k}
\;\neq\; 0
$$
for every $\lambda\neq 0$ and a.e. $E\in A$ w.r.t. the singular part of $\mu_{\tilde\omega,\Vv_\lambda,k}|A$.
This implies that the singular part of $\tilde \mu_\lambda|A$ vanishes for every $\lambda \neq 0$ and thus
the singular part of $\tilde \mu_\lambda$ vanishes also for Lebesgue a.e. $\lambda \in \RR$.

\vspace{.1cm}

It remains to consider the absolutely continuous part of $\tilde \mu_\lambda$.
Multiplying both sides of \eqref{eq-rank1pert5} with $|1+\lambda e_k^* \widehat G^z_\Vv e_k|^2$ and taking imaginary parts gives
\begin{eqnarray}
& & \!\!\!\!\!\!\!\!\! |1+\lambda \,e_k^* \widehat G^z_\Vv e_k|^2 \Im m (\Psi_1^* \widehat G^z_{\Vv_\lambda} \Psi_1)
\;=\;
|1+\lambda\, e_k^* \widehat G^z_\Vv e_k|^2\Im m \left(\Psi_1^* \widehat G^z_{\Vv} \Psi_1\right)
-\lambda\, \Im m \left(\Psi_1^* \widehat G^z_\Vv \Psi_k \Psi_k^* \widehat G^z_\Vv \Psi_1 \right)
\nonumber \\
&  & + \;\lambda^2\,\Big[ \Im m (e_k^* \widehat G^z_\Vv e_k)
\Re e(\Psi_1^* \widehat G^{z}_\Vv \Psi_k \Psi_k^* \widehat G^z_\Vv \Psi_1) \label{eq-rank1pert-poly}
-\Re e(e_k^* \widehat G^z_\Vv e_k) \Im m(\Psi_1^* \widehat G^{z}_\Vv \Psi_k \Psi_k^* \widehat G^z_\Vv \Psi_1)\Big]\;.
\end{eqnarray}
For $z\in\UM_1$, the r.h.s. of \eqref{eq-rank1pert-poly} is a second order polynomial in $\lambda$
which we denote by $P(z,\lambda)$. For $z=E+\imath\epsilon$ and $E\in A$, it converges as $\epsilon\downarrow 0$
to a limiting polynomial $P(E+\imath0,\lambda)$. As above consider
$$
A_{\Vv,k}\;=\;\left\{E\in\RR\,\left|\, \widehat G^{E+\imath0}_\Vv\;\text{exists and }\;\Im m(e_k^* \widehat G^{E+\imath 0}_\Vv e_k)>0\;\right. \right\}\;.
$$
{\bf Claim:} For $E\in A\cap A_{\Vv,k}$, $P(E+\imath 0,\lambda)$ cannot vanish identically as polynomial in $\lambda$.

\vspace{.1cm}

Suppose the contrary. Then by considering the constant and the linear term one deduces
$\Im m (\Psi_1^* \widehat G^{E+\imath0}_{\Vv_\lambda} \Psi_1)=0$ and
$\Im m \left(\Psi_1^* \widehat G^{E+\imath0}_\Vv \Psi_k \Psi_k^* \widehat G^{E+\imath0}_\Vv \Psi_1 \right)=0$.
Finally the quadratic term then gives
$\Im m(e_k^* \widehat G^{E+\imath 0}_\Vv e_k)\Re e\left(\Psi_1^* \widehat G^{E+\imath0}_\Vv \Psi_k \Psi_k^* \widehat G^{E+\imath0}_\Vv \Psi_1 \right)=0$.
As $E\in A_{\Vv,k}$, this now implies that one also has  $\Re e\left(\Psi_1^* \widehat G^{E+\imath0}_\Vv \Psi_k \Psi_k^* \widehat G^{E+\imath0}_\Vv \Psi_1 \right)=0$ so that $\Psi_1^* \widehat G^{E+\imath0}_\Vv \Psi_k \Psi_k^* \widehat G^{E+\imath0}_\Vv \Psi_1=0$. This is not the case
for $E \in A$ and hence the claim holds.

\vspace{.1cm}

Hence for $E\in A\cap A_{\Vv,k}$, $P(E+\imath0,\lambda)\neq 0$ for Lebesgue a.e. $\lambda\in\RR$.
As the set of $(E,\lambda)$ where this happens is clearly measurable, Fubini's theorem implies
that for Lebesgue a.e. $\lambda$ one has $P(E+\imath0,\lambda)\neq 0$ for Lebesgue a.e.
$E\in A\cap A_{\Vv,k}$. Since $|1+\lambda e_k^* \widehat G^{E+\imath 0}_\Vv e_k|^2$ exists and is strictly positive for
any $\lambda\in\RR$ and $E\in A\cap A_{\Vv,k}$, it follows from \eqref{eq-rank1pert-poly} that for a.e. $\lambda \in \RR$, Lebesgue a.e. $E\in A\cap A_{\Vv,k}$,
$\Im m(e_1^* \widehat G^{E+\imath 0}_{\Vv_\lambda} e_1)$ exists, is finite  and strictly positive.
Therefore for a.e. $\lambda\in\RR$, the absolutely continuous part of $\mu_{\tilde\omega,\Vv_\lambda,1}$ has almost surely a positive density on
$A\cap A_{\Vv,k}$. By Proposition~\ref{prop-rank1like}(i) the set $A_{\Vv_\lambda,k}$ coincides with $A_{\Vv,k}$
and, as $A$ has full Lebesgue measure, one obtains that $A\cap A_{\Vv,k}$ is an essential support of
$\mu_{\tilde\omega,\Vv_\lambda,k,{\rm ac}}$. Therefore for a.e. $\lambda \in \RR$, $\mu_{\tilde\omega,\Vv_\lambda,k,{\rm ac}}$ is
absolutely continuous w.r.t. $\mu_{\tilde\omega,\Vv_\lambda,1,{\rm ac}}$. This means that also the absolutely continuous part of $\tilde \mu_\lambda$ must vanish for a.e. $\lambda\in\RR$.
\hfill $\Box$

\vspace{.1cm}

\begin{coro}\label{cor-as-mu1}
For fixed $\tilde \omega$ and Lebesgue a.e. $\widehat \Vv \in \Jj  {\rm so}^*(2L)$,
the matrix valued measure $\mu_\omega$ is absolutely continuous
w.r.t. $\mu_{\omega,1}$. Hence for $\PP$ almost all $\omega=(\tilde\omega,\Vv)$
the measure $\mu_\omega$ is absolutely continuous w.r.t.
$\mu_{\omega,1}$.
\end{coro}

\noindent {\bf Proof.} Let $\omega$ be fixed.
By Proposition~\ref{prop-as-invert}, the assumptions of Theorem~\ref{theo-ac-relation} are fulfilled for a.e. $\widehat \Vv \in \Jj {\rm so}^*(2L)$. Therefore for a.e. $\tilde \Vv \in (\RR \Psi_k \Psi_k^*)^\perp$, the orthogonal complement of $\RR \Psi_k \Psi_k^*$ in $\Jj{\rm so}^*(2L)$, there is some $\lambda$ such that $\widehat \Vv_\lambda=\tilde \Vv+\lambda \Psi_k \Psi_k^*$ fulfills the assumptions of Theorem~\ref{theo-ac-relation}. Theorem~\ref{theo-ac-relation} now states, that for a.e. $\lambda \in \RR$, the measure $\mu_{\tilde\omega,\Vv_\lambda,k}$ is absolutely continuous w.r.t. $\mu_{\tilde\omega,\Vv_\lambda,1}$. For fixed $\tilde\omega$, the map $\widehat \Vv\mapsto (\mu_{\tilde\omega,\Vv,k},\mu_{\tilde\omega,\Vv,1})$ is Borelian as is the Lebesgue decomposition for finite measures which maps $(\mu,\nu)$ to the singular part of $\mu$ w.r.t. $\nu$. Hence the set of $\widehat \Vv$ where $\mu_{\tilde\omega,\Vv,k}$ is absolutely continuous w.r.t. $\mu_{\tilde\omega,\Vv,1}$ is measurable.
Therefore Fubini's theorem now implies that this set has full Lebesgue measure on $\Jj {\rm so}^*(2L)$.
This holds for any $k=2,\ldots,L$. As a finite intersection of sets of full measure is still a set of full measure we obtain that for a.e. $\widehat \Vv \in \Jj {\rm so}^*(2L)$ the measure $\sum_{k=1}^L \mu_{\tilde\omega,\Vv,k}$  is a.c. w.r.t. $\mu_{\tilde\omega,\Vv,1}$, namely $\mu_{\tilde\omega,\Vv}$ is a.c. w.r.t. $\mu_{\tilde\omega,\Vv,1}$.

\vspace{.1cm}

The maps $\omega\mapsto \mu_\omega$ and $\omega\mapsto \mu_{\omega,1}$ are Borelian. By the same arguments as above the set of $\omega=(\tilde \omega,\Vv)$ where $\mu_\omega$ is absolutely continuous w.r.t. $\mu_{\omega,1}$ is measurable. As the distribution $\pp_{\widehat \Vv} $ of $\widehat \Vv$ is absolutely continuous, we obtain that for any fixed $\tilde\omega$, for $\pp_\Vv$ almost every $\Vv$, $\mu_{\tilde\omega,\Vv}$ is a.c. w.r.t. $\mu_{\tilde\omega,\Vv,1}$.
By Fubini's theorem, we obtain that this is true for $\PP$ almost all $\omega$.
\hfill $\Box$

\vspace{.1cm}

\begin{theo}
\label{theo-acpure}
For $\PP$ almost every $\omega$ one has $\mu_{\omega,1,{\rm sing}}(\RR)=0$.
Together with {\rm Corollary~\ref{cor-as-mu1}} this implies
that for $\PP$ almost all $\omega$, one has $\mu_{\omega,{\rm sing}}(\RR)=0$.
\end{theo}

\noindent {\bf Proof.}
Let us define $A_\omega=\{E\,|\, \widehat G^{E+\imath 0}_\Vv\,\text{exists and}\; \Tr (\Im m(\widehat G^{E+\imath 0}_\Vv))> 0\}$ as well as $A_{\omega,k}=\{E\,|\,\widehat G^{E+\imath 0}_\Vv\;\text{exists and}\; \Im m(e_k^* \widehat G^{E+\imath 0}_\Vv e_k) > 0 \}$. By Lemma~\ref{lemma-start-id}(iii), one has $A_\omega = \bigcup_{k=1}^L A_{\omega,k}$. Clearly $A_\omega$ is an essential support of the a.c. part of $\mu_\omega$ and $A_{\omega,k}$  is an essential support of the a.c. part of $\mu_{\omega,k}$.

\vspace{.1cm}

By Kotani theory and Corollary~\ref{cor-as-mu1} for $\PP$ almost all $\omega$ the set $A_{\omega,k}$ has full Lebesgue measure and $\mu_\omega$ is a.c. w.r.t. $\mu_{\omega,1}$. Take such an $\omega=(\tilde\omega,\Vv)$. Then as $\mu_\omega$ is a.c. w.r.t. $\mu_{\omega,1}$ the sets $A_\omega$ and $A_{\omega,1}$ differ only by a set of measure zero and hence $\RR \setminus A_{\omega,1}$ is a set of zero Lebesgue measure. Let $\tilde {\Vv}$ be the projection of $\widehat \Vv$ orthogonal to $\Psi_1 \Psi_1^*$ and
$\pp_{\tilde \Vv}$ be the distribution of $\tilde \Vv$, namely the push forward of $\pp_{\widehat \Vv}$. Now set $\widehat \Vv_\lambda=\tilde \Vv + \lambda \Psi_1 \Psi_1^*$ and let $\Vv_\lambda$ be a pre-image of $\widehat{\Vv}_\lambda$ under the Cayley transformation. Then by Proposition~\ref{prop-rank1like} one has for
Lebesgue a.e. $\lambda \in \RR$, $\mu_{\omega_\lambda,1} (\RR \setminus A_{\omega_\lambda,1})=
\mu_{\omega_\lambda,1} (\RR\setminus A_{\omega,1})= 0$, where $\omega_\lambda=(\tilde\omega,\Vv_\lambda)$.
As $\mu_{\omega_\lambda,1,{\rm sing}}(A_{\omega_\lambda,1})=0$ by the definition of $A_{\omega_\lambda,1}$, this implies $\mu_{\omega_\lambda,1,{\rm sing}}(\RR)=0$. Now by Fubini's theorem for $\tilde\PP$ a.e. $\tilde\omega$ the situation described above happens for $\pp_\Vv$ a.e. $\Vv$. Then for $\pp_{\tilde \Vv}$ a.e. $\tilde \Vv$ we have $\mu_{\tilde\omega,\Vv_\lambda,1,{\rm sing}}(\RR)=0$ for Lebesgue a.e. $\lambda$.
Note that $\pp_{\tilde \Vv}$ is absolutely continuous and for fixed $\tilde \omega$ the set of $\Vv$ where
$\mu_{\tilde\omega,\Vv,1,{\rm sing}}(\RR)=0$ is measurable, because the map $\Vv\mapsto \mu_{\tilde\omega,\Vv,1}$ is Borelian as well as the Lebesgue decomposition. Fubini's theorem thus implies that for Lebesgue almost every $\widehat \Vv$ in the strip ${\rm supp}(\pp_{\tilde \Vv})+\RR \Psi_1\Psi_1^*$ one has $\mu_{\tilde \omega,\Vv,1,{\rm sing}}(\RR)=0$. As the distribution of $\widehat \Vv$ is supported in this strip, this also holds for $\pp_\Vv$ a.e. $\Vv$.

\vspace{.1cm}

As mentioned, this situation happens to be true for $\tilde \PP$ a.e. $\tilde \omega$. By the same arguments as above the set of $\omega$ where $\mu_{\omega,1,{\rm sing}}(\RR)=0$ is measurable. Fubini's theorem now implies that $\mu_{\omega,1,{\rm sing}}(\RR)=0$ for $\PP$ a.e. $\omega$. Since for $\PP$ a.e. $\omega$ one also has that $\mu_{\omega}$ is a.c. w.r.t. $\mu_{\omega,1}$, we finally obtain that $\mu_{\omega,{\rm sing}}(\RR)=0$ for $\PP$ a.e. $\omega$.
\hfill $\Box$

\vspace{.2cm}

\noindent {\bf Proof of second claim of Theorem~\ref{theo-main}(ii).} This
is Theorem~\ref{theo-acpure}. 
\hfill $\Box$



\begin{thebibliography}{99}

\bibitem[BL]{BL} P. Bougerol, J. Lacroix, {\sl Products of Random
Matrices with Applications to Schr{\"o}dinger Operators}, (Birkh{\"a}user,
Boston, 1985).

\bibitem[Bou]{Bou} H. Boumaza,
{\sl Localization for a matrix-valued Anderson model}, preprint
arXiv:0902.1628.
 
\bibitem[BP]{BP}
S. de~Bievre, J.~V. Pul\'e,
{\sl Propagating Edge States for a Magnetic Hamiltonian},
Math. Phys. Elect. Journal {\bf 5}, 17 pages (1999).

\bibitem[EM]{EM} F. Evers, A.~D. Mirlin, {\sl Anderson transitions},
Rev. Mod. Phys. {\bf 80}, 1355 (2008).

\bibitem[FGW]{FGW} J. Fr\"ohlich, G. M. Graf, J. Walcher,
{\sl On the extended nature of edge states of Quantum Hall
Hamiltonians}, Ann. H. Poincar\'e {\bf 1}, (2000).

\bibitem[GT]{GT} F. Gesztesy, E. Tsekanovskii, {\sl On matrix-valued Herglotz
functions}, Math. Nachr. {\bf 218}, 61-138 (2000).

\bibitem[GM]{MG} I. Goldsheid, G. Margulis, {\sl Lyapunov indices of a product of
random matrices}, Russ. Math. Surveys {\bf 44}, 11-71 (1989).


\bibitem[GR]{GR} Y. Guivarch, A. Raugi, {\sl Froni\`ere de
  Furstenberg, propri\'et\'es de contraction et th\'eor\`emes de
  convergence}, Z. f. Wahrscheinlichkeitstheorie verw. G. {\bf 69},
  187-242 (1987). 

\bibitem[HS]{HS} D.~B. Hinton, A. Schneider,
{\sl On the Titchmarsh-Weyl coefficients for singular S-Hermitian Systems I},
Math. Nachr. {\bf 163}, 323-342 (1993).



\bibitem[JL]{JL} V. Jaksic, Y. Last, {\sl Spectral structure of Anderson type Hamiltonians},
Inventiones Mathematicae {\bf 141}, 561-577 (2000).

\bibitem[JSS]{JSS} S. Jitormiskaya, H. Schulz-Baldes, G. Stolz,
{\sl Delocalization in random polymer models},
Commun. Math. Phys. {\bf 233}, 27-48 (2003).

\bibitem[KRS]{KRS} J. Kellendonk, Th. Richter, H. Schulz-Baldes,
{\sl Edge channels and Chern numbers in the integer quantum Hall effect},
Rev. Math. Phys. {\bf 14}, 87-119 (2002).

\bibitem[KLS]{KLS} A. Klein, J. Lacroix, A. Speis, {\sl Localization
for the Anderson model on a strip with singular potentials},
J. Funct. Anal. {\bf 94}, 135-155 (1990).

\bibitem[KS]{KS} S. Kotani, B. Simon, {\sl Stochastic Schr\"odinger
Operators and Jacobi Matrices on the Strip},
Commun. Math. Phys. {\bf 119}, 403-429 (1988).

\bibitem[LM]{LM} M. Lesch, M. Malamud, {\sl On the number of square integrable solutions
and self-adjointness of symmetric first order systems of differential equations},
J. Diff. Equ. {\bf 189}, 556-615 (2003).

\bibitem[SB1]{SB1} H. Schulz-Baldes,
{\sl Rotation numbers for Jacobi matrices with matrix entries}, Math. Phys.
Elect. Journal {\bf 13}, 40 pages (2007).

\bibitem[SB2]{SB2} H. Schulz-Baldes,
{\sl Geometry of Weyl theory for Jacobi matrices with matrix entries}, preprint
2008, to appear in J. d'Analyse Math\'ematique.

\bibitem[Sun]{Sun} F. Sun, {\sl Kotani theory for stochastic Dirac operators},
Northeast. Math. J {\bf 9}, 49-62 (1993).


\end{thebibliography}
\end{document}